\documentclass{aastex62}

\def\las{\mathrel{\hbox{\rlap{\hbox{\lower3pt\hbox{$\sim$}}}\hbox{\raise2pt\hbox
{$<$}}}}}
\def\gas{\mathrel{\hbox{\rlap{\hbox{\lower3pt\hbox{$\sim$}}}\hbox{\raise2pt\hbox
{$>$}}}}}
\newcommand{\degree}{^{\circ}}

\received{2019 December 19}
\revised{2020 February 6}
\accepted{2020 February 8}
\submitjournal{Planetary Science Journal}

\shorttitle{DART-ejected meteoroids}
\shortauthors{Wiegert}

\begin{document}

\title{On the delivery of DART-ejected material from asteroid (65803) Didymos to Earth}

\correspondingauthor{Paul Wiegert}
\email{pwiegert@uwo.ca}

\author[0000-0002-1914-5352]{Paul Wiegert}
\affiliation{Dept. of Physics and Astronomy, The University of Western Ontario, London Canada}
\affiliation{Institute for Earth and Space Exploration (IESX), The University of Western Ontario, London Canada}



\begin{abstract}

The DART spacecraft is planned to impact the secondary of the binary
asteroid (65803) Didymos in 2022, to assess deflection strategies for
planetary defense. The impact will create a crater and release
asteroidal material, some of which will escape the Didymos system.
Because the closest point of approach of Didymos to Earth's orbit is
only 6 million km (about 16 times the Earth-Moon distance), some
ejected material will make its way sooner or later to our planet, and
the observation of these particles as meteors would increase the
scientific payout of the DART mission. The DART project may also
represent the first human-generated meteoroids to reach Earth, and
a test case for human activity on asteroids and
its eventual contribution to the meteoroid environment and spacecraft
impact risk.  This study examines the amount and timing of the
delivery of meteoroids from Didymos to near-Earth space.

This study finds that very little DART-ejected material will reach our
planet, and most of that only after thousands of years. But some
material ejected at the highest velocities could be delivered to
Earth-crossing trajectories almost immediately, though at very low
fluxes. Timing and radiant directions for material reaching Earth
are calculated, though the detection of substantial numbers would
indicate more abundant and/or faster ejecta than is expected.

The DART impact will create a new meteoroid stream, though probably
not a very dense one. However, larger, more capable asteroid impactors
could create meteoroid streams in which the particle flux exceeds that
naturally occurring in the Solar System, with implications for
spacecraft safety.
  
\end{abstract}



\section{Introduction} \label{sec:intro}

The Asteroid Impact \& Deflection Assessment (AIDA) mission is a
partnership between NASA and ESA. The Double Asteroid Redirection Test
(DART, spearheaded by NASA) will send a spacecraft to impact the
secondary of the (65803) Didymos binary asteroid system, in order to
determine the efficiency of kinetic impactors as a strategy for asteroid
deflection. The Hera spacecraft (led by ESA) is planned to observe
and characterize the impact's effects on the Didymos system
\citep{micchekup16,cherivmic18}.

Envisioned as a planetary defense exercise, the project may also
produce the first artificially-generated meteoroids that reach
Earth. The Deep Impact spacecraft that struck comet 9P/Tempel 1 in
2005 \citep{ahebeldel05} would have released a similar cloud of
material but its Minimum Orbital Intersection Distance (or MOID) with
Earth is over 0.5~AU, a order of magnitude further than Didymos' MOID
of 0.04~AU, so material is not delivered as efficiently to our planet.
The observation of DART-generated meteors if and when they reach us
would provide additional information about the target and
impact-related processes. The DART impact is also a test case in some
sense for the production of asteroid-derived debris by human activity
such as asteroid mining, though mining is perhaps more likely to
release material at lower speeds (e.g. \cite{flabolbye19}). At the
very least, it seems prudent to ask how much material might be
delivered to near-Earth space as a result of such an impact, and that
is the purpose of this study.

\subsection{Didymos' current orbit}

The point of closest approach of Didymos' orbit to Earth's orbit (its
Minimum Orbital Intersection Distance or MOID) sits currently at
0.0398~AU. The Jet Propulsion Laboratory (JPL)'s Solar System Dynamics
site\footnote{https://ssd.jpl.nasa.gov/sbdb.cgi?sstr=Didymos;
  retrieved 5 Dec 2019} lists it as a near-Earth object (NEO,
indicating that its perihelion distance lies within 1.3 AU of the Sun)
and a Potentially Hazardous Asteroid (PHA, indicating that it poses at
least a hypothetical danger of impact, having an Earth MOID less than
0.05 AU and an absolute magnitude $H$ less than 22.0
\citep{atkticwil00,stoyeobot03})\footnote{The absolute magnitude provides a
measure of the asteroid size if its albedo is known. Because absolute
magnitude increases as asteroid size decreases, the $H<22$ limit
corresponds to a minimum asteroid diameter of between 100 and 250
meters, for typical asteroidal albedos ranging from 0.25 to 0.05,
respectively.}. The Didymos primary itself has $H=18.2$ and a diameter of 780
meters \citep{micchekup16}.

The MOID of Didymos is currently slowly decreasing over time under the
gravitational perturbations of the other planets of our Solar System,
but it will reach a minimum value of 0.022~AU in 2500 years and then will
start to increase. As a result, Didymos is not considered to be an
impact threat to Earth in the near-term and has a Torino scale
\citep{bin00,morchaste04} rating of
zero\footnote{https://cneos.jpl.nasa.gov/sentry/, retrieved 12
  December 2019}.

The Didymos asteroid pair consists of a primary with a diameter of
780~m, and a 163~m diameter secondary; their individual centers of
mass are separated by 1.18 km and their orbital period is 11.92 hours
\citep{micchekup16}. Material released from Didymos' secondary
(informally known as 'Didymoon' e.g. \cite{micchekup16}) by the DART
impactor can be expected to show a range of behaviors.  Ejecta with
speeds less than the escape speed from Didymos (about 1 m/s,
\cite{micchekup16}) can be expected to remain in orbit around the
asteroid and/or re-impact one of the members of the binary.  Ejected
material that remains bound to the binary system may prove hazardous
to the follow-up spacecraft Hera slated to observe the impact and has
been examined by other authors \citep{ricobr16,micyu17}. However, this
material cannot reach Earth and won't be considered here.

Ejecta released at speeds at or above 1 m/s will mostly escape from
the vicinity of Didymos and its moon, and begin to orbit the
Sun. Slower material will follow a path very close to that of Didymos
itself, either leading or trailing that asteroid.  This material will
have a MOID similar to Didymos' and so will not be able to reach Earth
(at least not immediately, see Section~\ref{sec:results}). Material
ejected at higher speeds will travel on orbits further from
Didymos'. These particles may have MOIDs closer to Earth, and could
potentially reach it. An additional complication is that for small
particles (less than about 1 cm in size), radiation pressure and
Poynting-Robertson drag will affect the delivery of debris to our
planet.

Particles ejected from Didymos cannot impact Earth unless their MOID
decreases to a sufficiently small value: for this study we will
require that the MOID reach zero for an impact with Earth to be
considered possible. When we are considering delivery of material to
near-Earth space however, larger MOIDs will be of interest. For
example, the Planck spacecraft \citep{tau04} currently orbits near,
and the James Webb Space Telescope (JWST \citep{garmatcla06}) will
soon be launched to, the Earth-Sun L2 point, which is about 0.01~AU
outside Earth's orbit. The SOHO spacecraft moves near the inner
Earth-Sun L1 point \citep{domflepol95}, located a similar distance
inside our planet's orbit. Because these important space assets are
vulnerable in principle to meteoroid impacts, we will use a MOID of
0.01~AU with Earth as our boundaries for 'near-Earth space'. Of course,
a particle with a low or zero MOID with Earth will still not
necessarily impact our planet: both planet and particle must
be at the point where their orbits cross at the same time and this
must be checked for as well.

\subsection{Ejecta} \label{sec:ejecta}

The DART impact is expected to produce a final crater of size around
10m in diameter on Didymoon with $10^4 - 10^5$ kg of mass escaping
($\sim 0.01$\% of the secondary's mass, \cite{stiatcbar15}). Larger
ejected masses and larger craters (even exceeding 100~m diameter on the
163~m diameter moon) are possible though less likely, and the precise
outcome depends on the unknown strength, porosity and other physical
properties of Didymoon. Other Didymoon-specific simulations of the
impact show ejecta speeds approaching the impactor speed (expected to
be 6 km/s, \cite{cherivmic18}) in some cases, though the bulk of
material is released at much lower (several to hundreds of m/s) speeds
\citep{ricobr16, raddavcol19, stisyache20}.  We note that the
cratering produced by Deep Impact proved difficult to interpret.
Fast-moving ejecta implied acceleration due to expanding gases,
probably from volatiles \citep{holhou07} that are less likely to be
present in Didymoon. As a result, the Deep Impact experiment does not
provide strong constraints on the ejecta to be expected from DART.

Here we will examine three specific ejection speeds, 10 m/s, 100 m/s
and 1000 m/s. These are chosen to span the range between the lowest
speeds that can escape ($\gas~1$~m/s) and the highest expected
ejection speeds. These speeds reach somewhat higher than typical of
the cometary ejection processes (tens to hundreds of m/s, \cite{whi51,
  jon95, cri95}) which create most meteoroid streams.

Particle sizes examined in this study have diameters of 10~$\mu$m,
100~$\mu$m, 1~mm, and 1 cm. For a traditional power-law size
distribution larger particles would be less common, but it is not
clear if that is a good assumption for Didymoon. Some asteroids, such
as (25143) Itokawa, (162173) Ryugu and (101955) Bennu, have abundant
cm and larger sized particles on their surface \citep{fujkawyeo06,
  wathirhir19, laudelben19}. On the other hand, the spinning-top shape
of the Didymos primary \citep{cherivmic18} indicates Didymoon may be
made of particles lifted from it by centrifugal forces, a process
which could favour fine particles. We do not assume a particular size
distribution here. Our largest particle size of 1 cm is chosen because
particles larger than this will evolve dynamically in the same way, as
radiation effects are negligible at these sizes on the time scales
considered here. Particles smaller than 10~$\mu$m may be produced in
substantial quantities, but they are not modeled here because they
almost undetectable with current meteor techniques. Optical meteor
systems typically see millimeter-sized particles, while meteor patrol
radars like the Canadian Meteor Orbit Radar (CMOR, \cite{jonbroell05})
can typically see particles down to masses of $10^{-8}$~kg \citep{blacamwer11} which
corresponds to a 200 $\mu$m diameter at our assumed density. Particles
smaller than $10~\mu$m are also well below the usual threat limit to
spacecraft set by NASA's Meteoroid Environment Office (MEO) of
$10^{-9}$~kg or 100~$\mu$m diameter at our assumed density
\citep{mookoecoo15}.


As for the DART impact timing, that is not yet set. Planned for 'late
September' \citep{micchekup16}, more recent work \citep{cherivmic18}
lists `5 October'. The impact will most likely occur near Didymos' closest
approach to Earth, which the JPL website gives as 2022 October 4.
Here we will take our baseline scenario to be an impact date of 2022
October 1 at 0000 UT but will examine other dates as well.

\section{Methods}

The dynamics of the particles were simulated with the RADAU15
\citep{eve85} algorithm with an accuracy parameter of $10^{-12}$.  The
baseline impact date calculations were verified by repeating them with
the Wisdom-Holman \citep{wishol91} algorithm modified to handle close
approaches by the hybrid method \cite{cha99}, with a time step of 2
days or less. The differences between the two were negligible and we
report on the RADAU15 results here. The particles were simulated in a
solar system which includes the Sun and all eight planets (the Moon
was not simulated independently but its mass was included at the
Earth-Moon barycenter) with their initial positions derived from the
JPL DE405 ephemeris. The orbit of Didymos was obtained from the JPL
SSD website\footnote{https://ssd.jpl.nasa.gov/sbdb.cgi, retrieved 2019
  November 2}. These orbital elements (see Table~1) were advanced to the
desired date of the impact to be simulated.

\begin{table}
\centering
\begin{tabular}{lc}\hline \hline
  $a$ (AU)  & 1.644267944704789 \\
  $e$       & 0.3840204904781526 \\
  $i$ ($\degree$) &  $3.408561576582074$ \\
  $\Omega$ ($\degree$) &  73.20707875073758 \\
  $\omega$ ($\degree$) & 319.3188820727824 \\
  $M$ ($\degree$) &  124.6176912105528\\ \hline
\hline
\end{tabular}
\caption{The orbital elements (J2000) of Didymos from JPL Horizons. The semimajor axis $a$, eccentricity $e$, inclination $i$, longitude of the ascending node $\Omega$, argument of perihelion $\omega$, and mean anomaly $M$ at 2458600.5 (2019-Apr-27.0) TDB. \label{ta:elems}}
\end{table}

Ejecta were modeled as massless particles released from a spherical
shell located 1 km from Didymos with relative velocity directions
chosen randomly on the sphere. Though the release of ejecta will
primarily be in a cone aligned along the impact vector rather than
spherical, that impact vector is not yet known. Our choice encompasses
all impact geometries for simplicity and completeness. However, the
impact is planned for the Earth-facing side and this will increase the
flux of particles delivered directly to us (see
Section~\ref{sec:direct}), though impact location is not expected to
have significant effect on the long-term evolution of the meteoroid
stream.  DART's target Didymoon orbits 1.18 km from the center of the
primary, which is why the 1 km release distance was chosen, though
this offset has little effect. The orbital speed of Didymoon ($\approx
0.2$ m/s) is negligible for our simulations, as is the gravity of both
asteroid components, and these are ignored. One thousand particles are
released for each size/impact date/ejection speed combination.

The effect of radiation pressure and Poynting-Robertson drag are
included.  Each particle is assigned diameter $d$ which defines its
$\beta$ parameter (the ratio of radiation to gravitational forces)
according to an expression derived from that of \cite{weijac93}
\[
\beta = \frac{1.14 \times 10^{-3}}{\rho d}
\]
where $d$ is the particle diameter in meters, and $\rho$ is the
assumed density in kg~m$^{-3}$ which we take to be the bulk density of
the Didymos system, 2100 kg/m$^{-3}$ \citep{micchekup16,
  cherivmic18}. Since Didymos is an Sq spectral type, associated with
ordinary chondrite meteorites (densities of 3600-3900 kg/m$^3$
\cite{briyeohou02}), our choice of density may be low. If this is the
case, the dynamics of the simulations presented here all remain
correct, except that the particles correspond to somewhat different
sizes or masses. For example, if the actual particle density was
double that assumed, the mass quoted for a particular size particle
would have to be increased by a factor of 2, or the diameter for a
given mass particle increased by a factor of $2^{1/3} \approx 1.3$.

Because the results we found to be somewhat sensitive to the timing of the
impact, we examined impacts on 2022 October 1 (baseline), then at
intervals of $\pm$ 1 month, $\pm 3$ months, and then an impact
at the asteroid's aphelion (2023 November 10)

\section{Results and discussion} \label{sec:results}

\begin{figure*}
\gridline{\fig{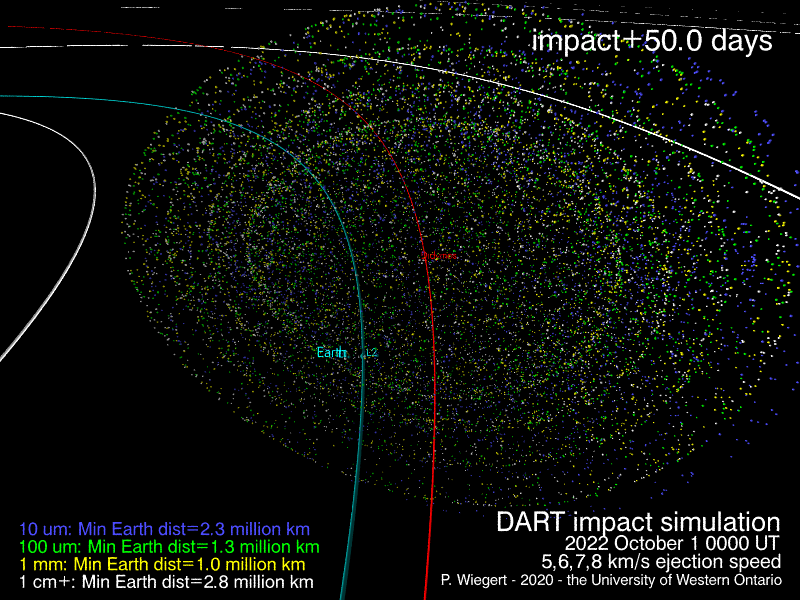}{0.5\textwidth}{(a)}
          \fig{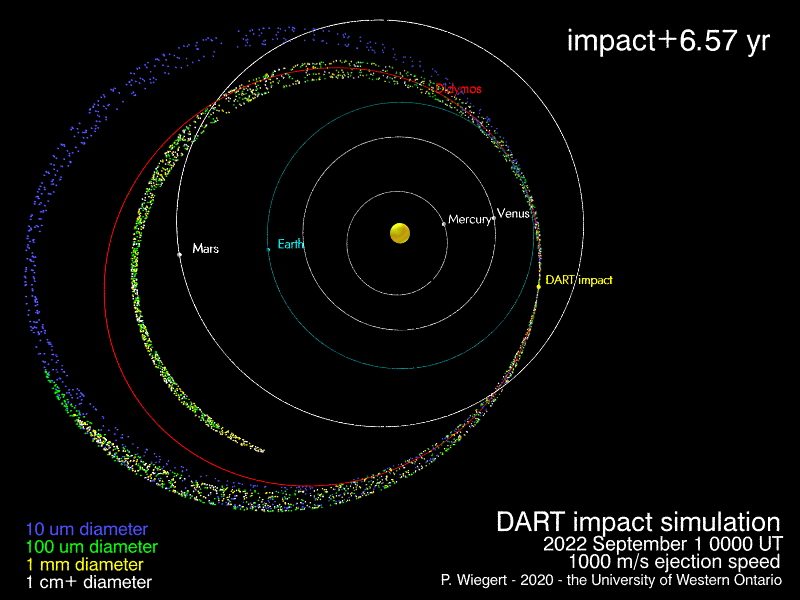}{0.5\textwidth}{(b)}
          }
\caption{(Animated figure). Panel a: The dispersal of ejecta at 5, 6,
  7 and 8 km/s from a 2022 October 1 impact.  Panel b: The dispersal
  of ejecta at 1000 m/s from a 2022 September 1 impact. See the text
  for more details. [These animations are stored permanently at the Planetary Science Journal and semi-permanently at 
    \url{http://www.astro.uwo.ca/~wiegert/Didymos/Didymos-DART-animA.mp4}
    and
    \url{http://www.astro.uwo.ca/~wiegert/Didymos/Didymos-DART-animB.mp4}. ] \label{fig:animated}}
\end{figure*}

The dispersion time $T$ for the material to spread around the mean
orbit depends primarily on the ejection speed. For 1000 m/s
ejection, $T \las 10$~yr (see e.g. Fig.~\ref{fig:animated}b),
for 100 m/s ejection, $20 <T <200$~yr, and for 10 m/s ejection, $150 <
T <500$ yrs.  Since the dispersion timescale is much shorter than the
time it takes for the particle MOID to get close to Earth in most
cases, we can estimate the delivery time for meteoroids to near-Earth
space as being the time it takes for the MOID to drop to a suitable
value. Those cases where the ejecta is not fully dispersed will be
called out specifically when they arise.

\begin{figure*}
\gridline{\fig{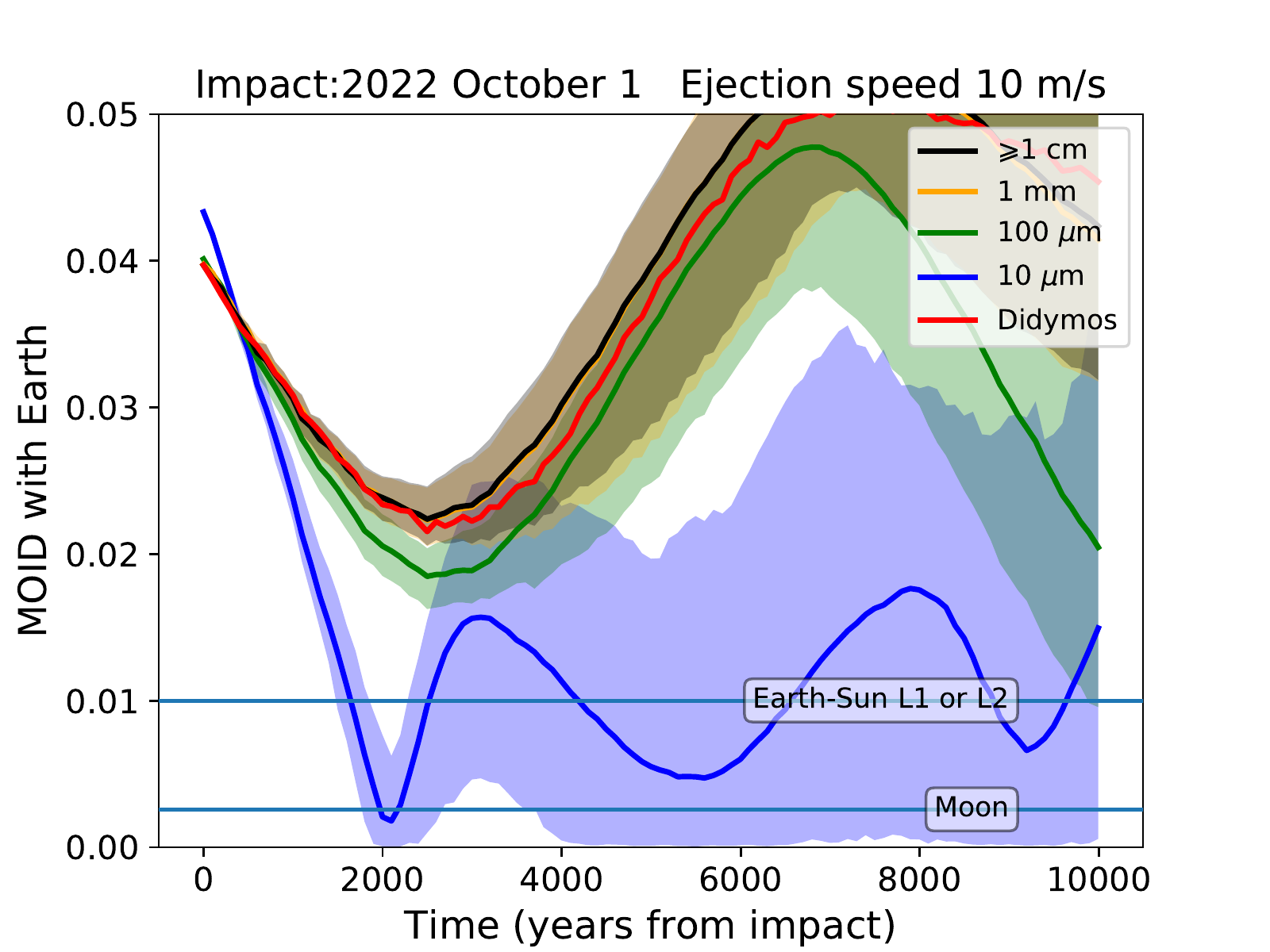}{0.33\textwidth}{(a)}
          \fig{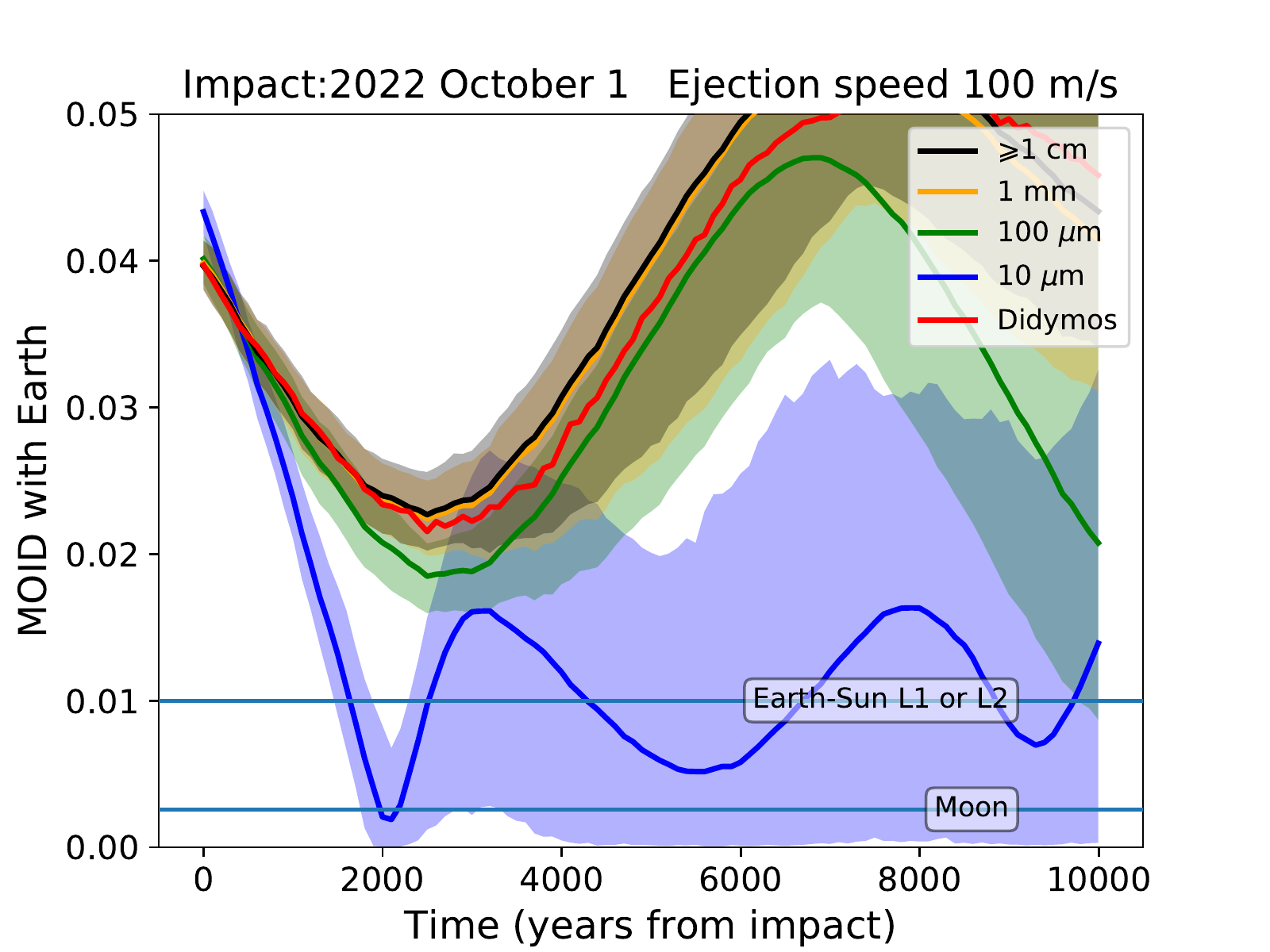}{0.33\textwidth}{(b)}
          \fig{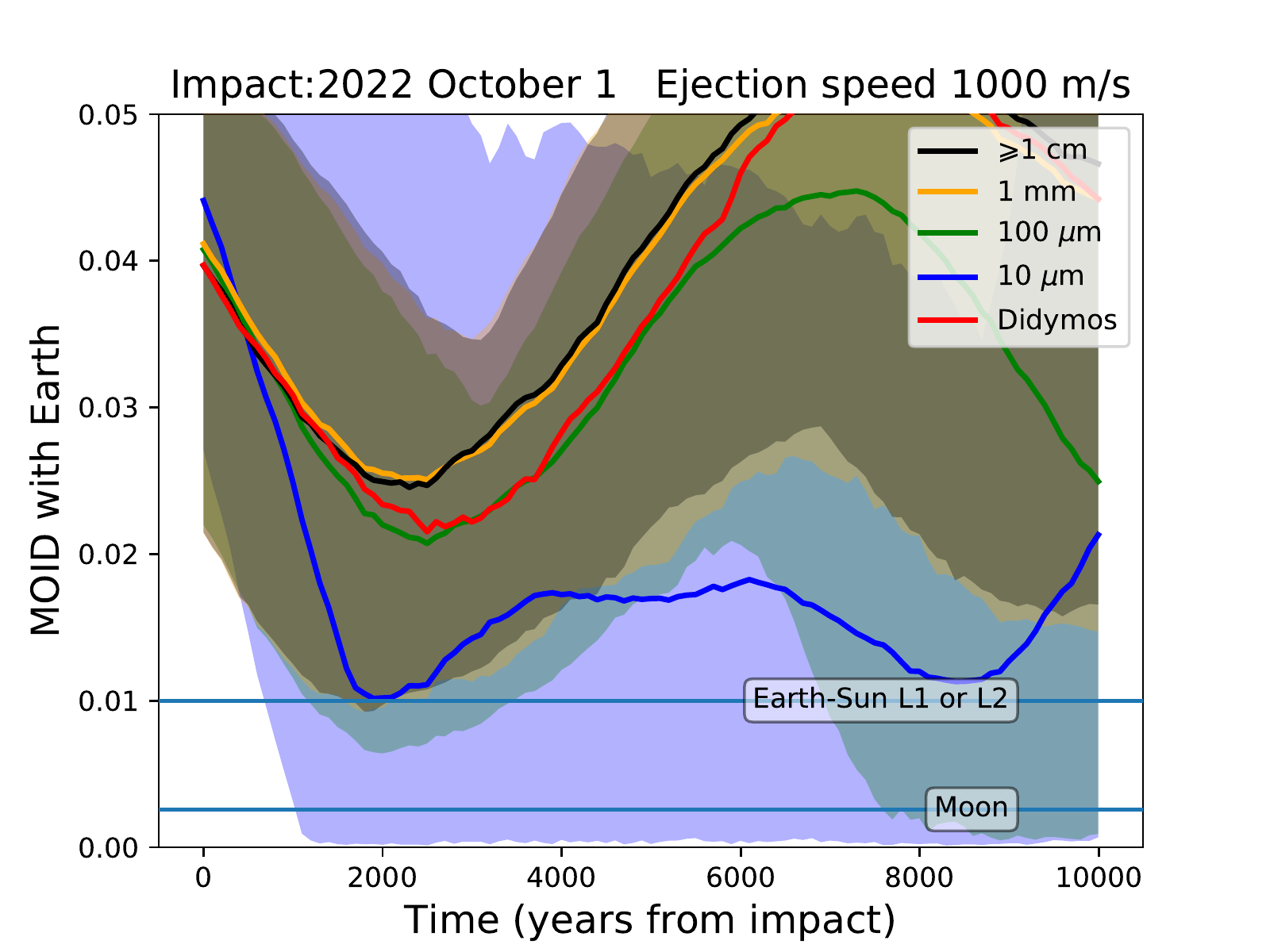}{0.33\textwidth}{(c)}
          }
\caption{The evolution of the MOIDs of the simulated DART-ejected
  debris in the case of an impact on 2022 October 1. The median value
  of the MOIDs is shown by a solid line, while the shaded areas
  indicate the highest and lowest values. The MOID of Didymos itself
  is shown in red.\label{fig:1Oct2022}}
\end{figure*}

The time evolution of the MOIDs for the baseline case is shown in
Figure~\ref{fig:1Oct2022}. The figure includes the lowest and highest
MOID values, or rather we will take the 'lowest' and 'highest' values
here to mean the 1$^{st}$ and $99^{th}$ percentile values. This
excludes the 10 truly lowest and highest MOIDs of the 1000 simulated
particles in each case. This choice is made because, while a spherical
distribution of ejecta velocities produces a meteoroid stream with a
well-defined elliptical cross-section (see Fig.~\ref{fig:animated}a
later), over time a few particles are scattered widely by close
planetary encounters, and their inclusion would produce deceptive
results. Here we are interested in when the bulk of the ejected
population might reach Earth, not rare heavily-perturbed particles.

From Figure~\ref{fig:1Oct2022} it is clear that for particles with the
range of speeds and sizes chosen, none are delivered immediately to
Earth and many of them will never reach us (at least not on the 10,000
year time scale examined here) though this is not true of all impact
times. The smallest simulated particles (10~$\mu$m) reach us first,
after 1000-2000 years. Their different evolution is a result of the
effects of radiation forces, which affect the orbits of these small
particles strongly immediately upon release, and they do not follow
the orbit of their parent asteroid as closely as larger particles. In
particular, their MOIDs immediately after release differ from those of
other particles because radiation pressure counteracts solar
gravity substantially as particle size diminishes. Effectively,
radiation pressure's radial outward force means that small particles
behave as if the Sun's mass were reduced, and thus they have larger
orbits for the same initial velocity (see e.g. Animated
Figure~\ref{fig:animated}b). As a result, particles' with different
sizes may have different initial MOIDs, even though they are all on
orbits released from the same point at the same speed.

Larger particles are much less affected by radiation pressure, closely
follow the dynamical evolution of Didymos itself and don't reach
near-Earth space. The exception is for the highest ejection speeds
simulated, 1000 m/s. Here the orbital dispersion caused by the higher
ejection speed brings meteoroids of all sizes just barely into
near-Earth space in 1500-2000 years (Figure~\ref{fig:1Oct2022}c).

\begin{figure*}
\gridline{\fig{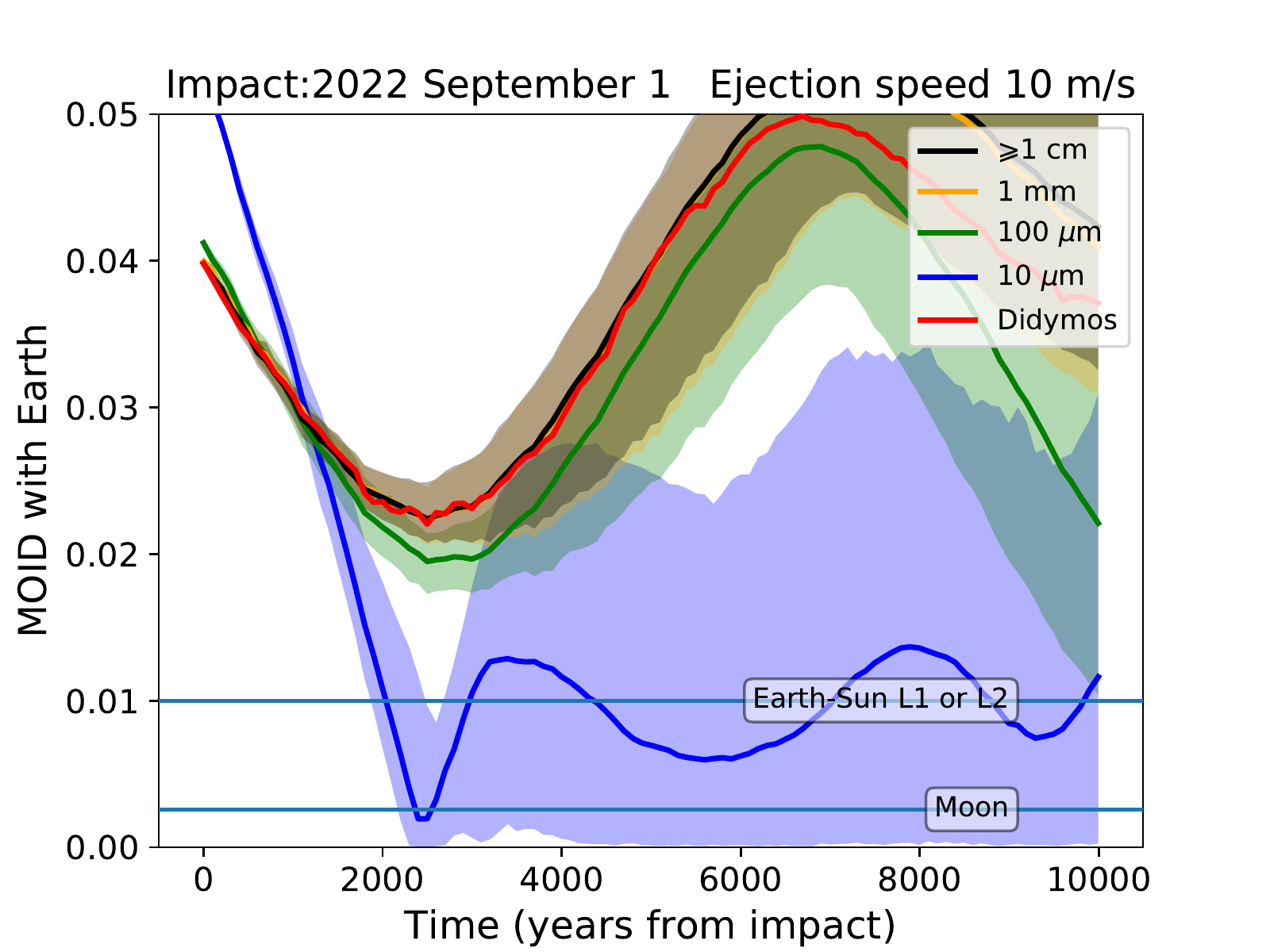}{0.33\textwidth}{(a)}
          \fig{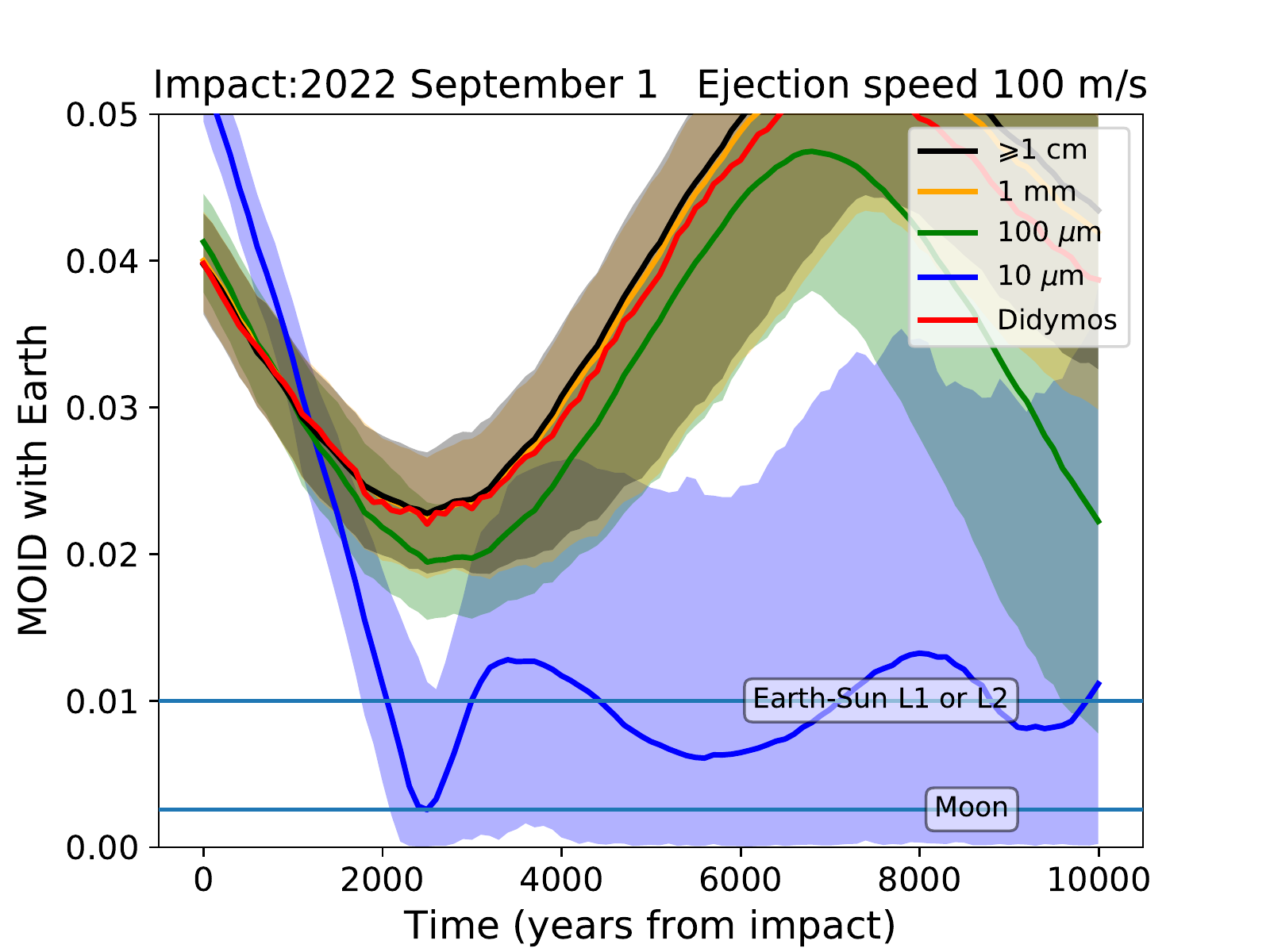}{0.33\textwidth}{(b)}
          \fig{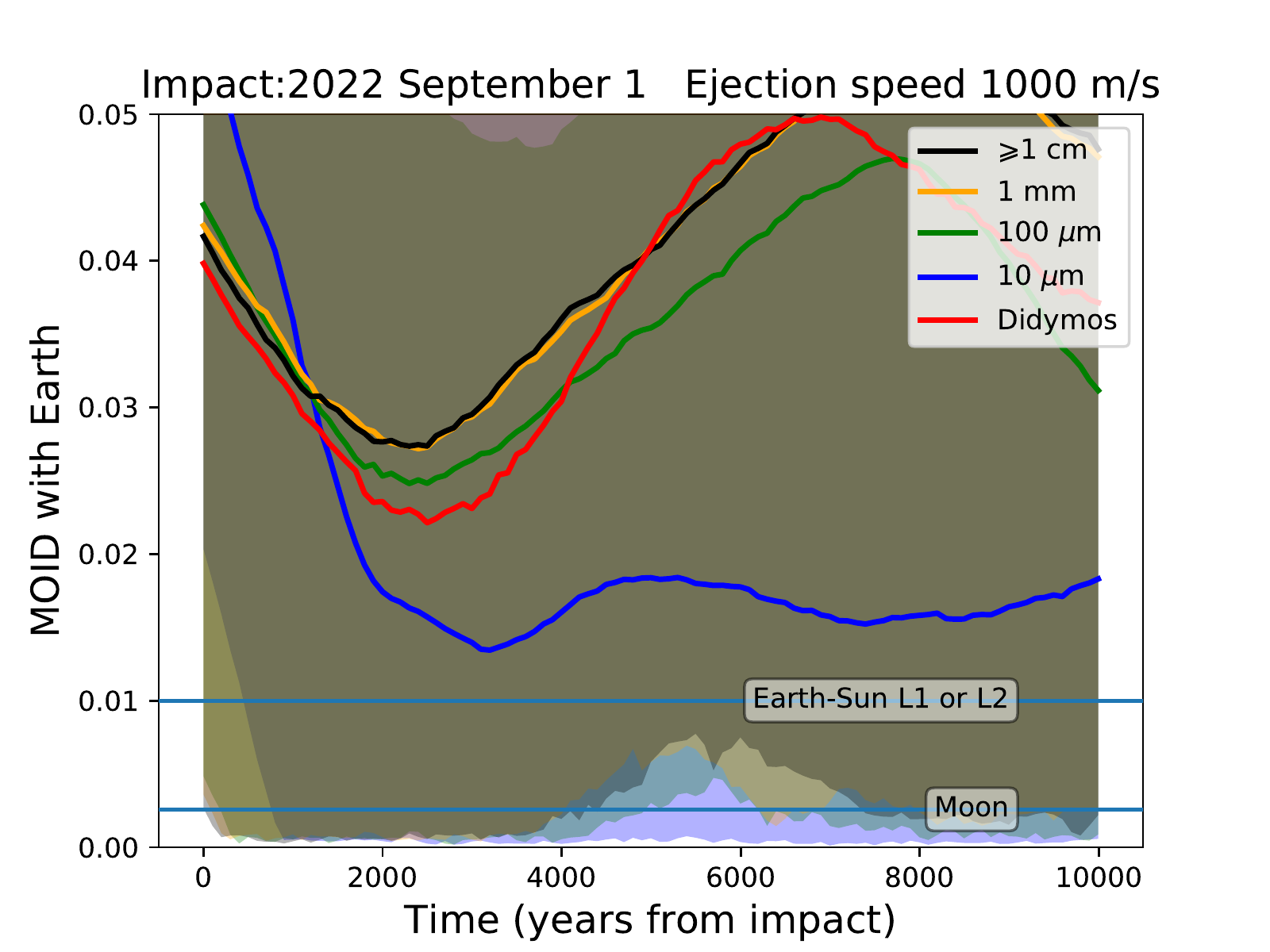}{0.33\textwidth}{(c)}
          }
\gridline{\fig{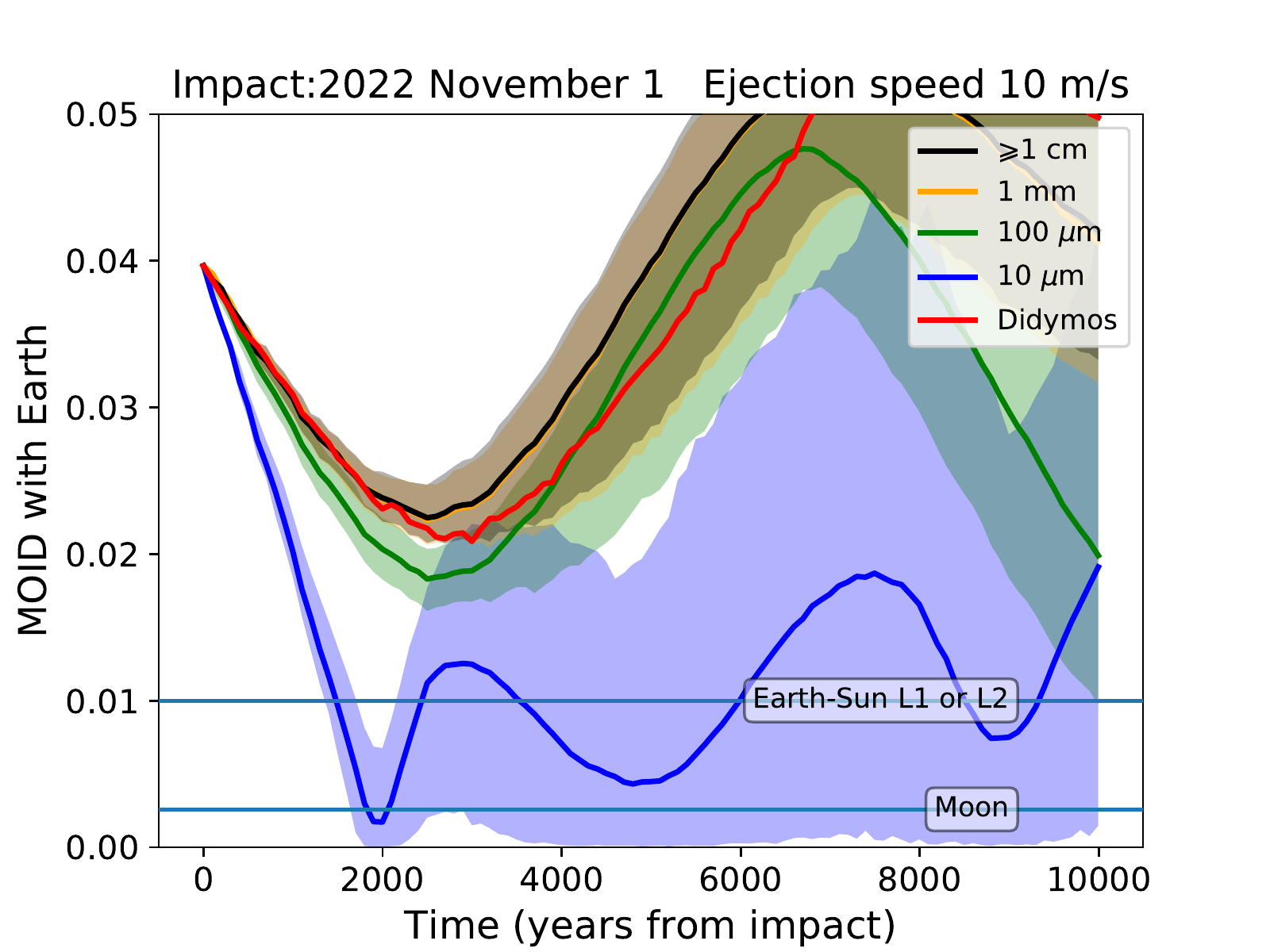}{0.33\textwidth}{(a)}
          \fig{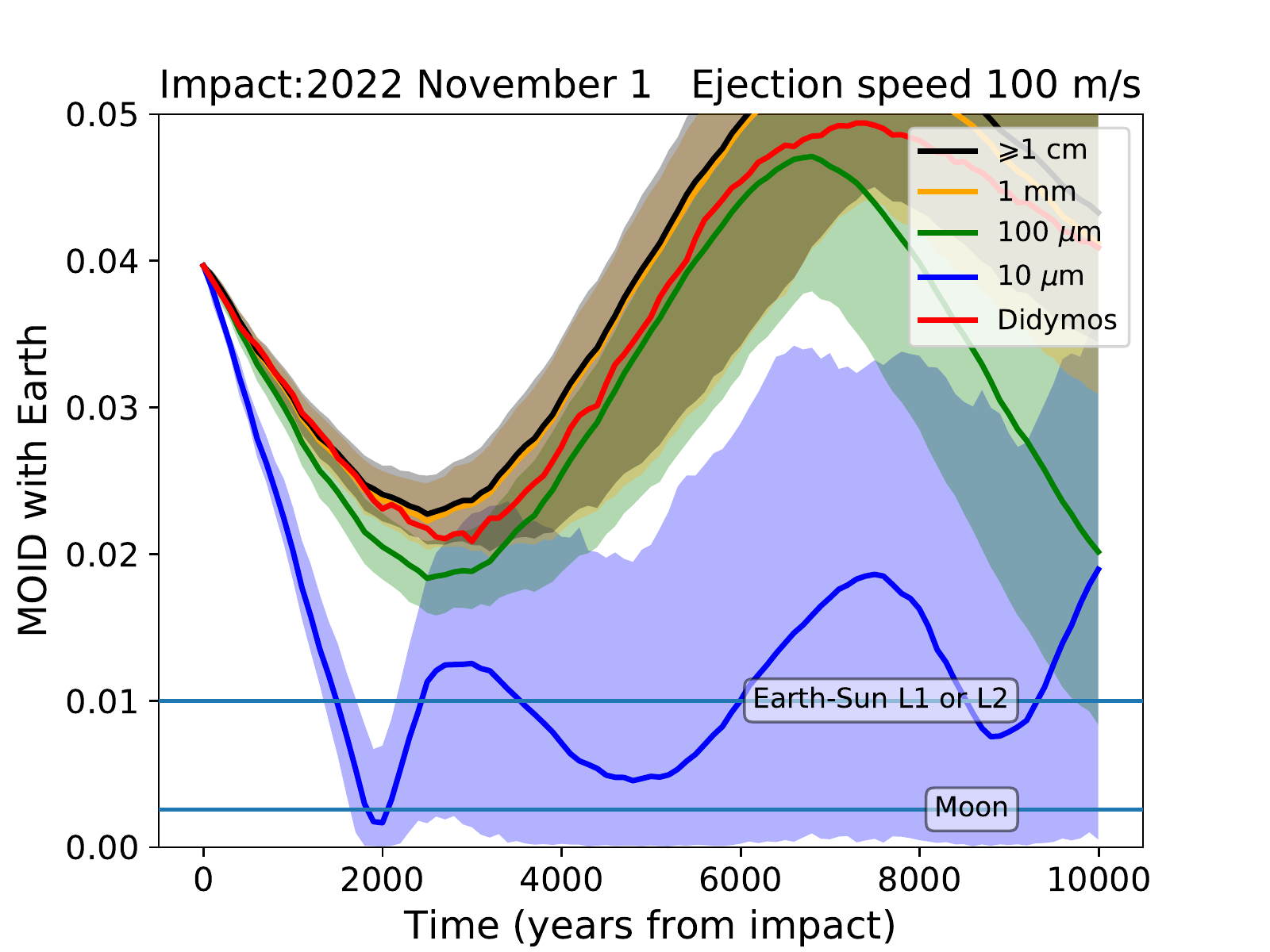}{0.33\textwidth}{(b)}
          \fig{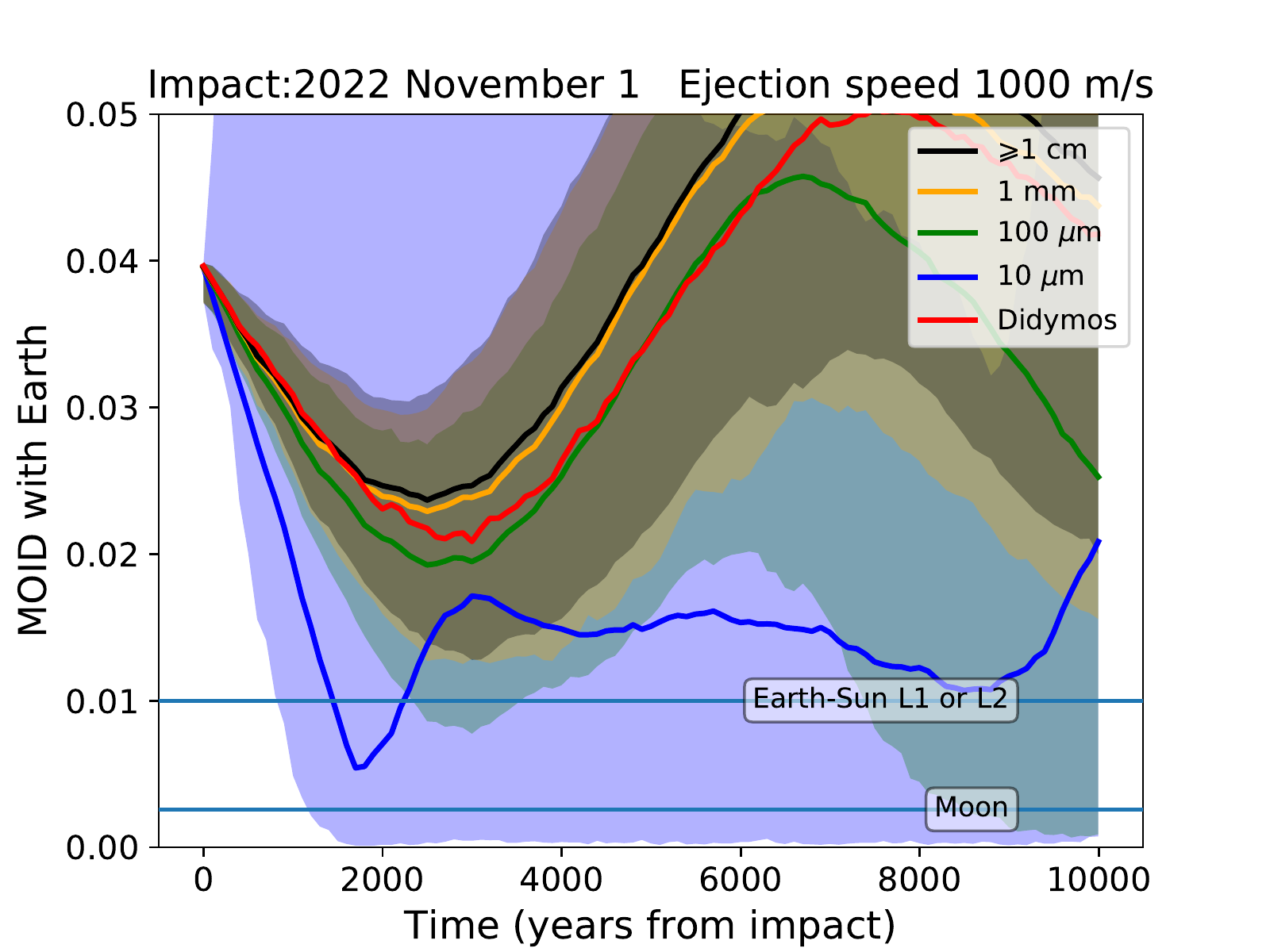}{0.33\textwidth}{(c)}
          }
\caption{The evolution of the MOIDs of the simulated DART-ejected debris in the case of an impact on 2022 September 1 or 2022 November 1. See Figure~\ref{fig:1Oct2022} for more details.\label{fig:1SepNov2022}}
\end{figure*}

Figure~\ref{fig:1SepNov2022} shows the effect of a change in time of
the impact by one month. The overall result is much like that of the
baseline impact date. Our smallest 10$\mu$m particles arrive in about
2000 years, larger ones are delayed or do not reach near-Earth space
at all on these time frames. The only exception is the case of the
highest ejection speeds for an impact on 2022 September 1. Here the
particles are widely enough dispersed to have MOIDs intersecting 
Earth almost immediately.

We also examined the case of the impact occurring three months early or
later, or at the asteroid's aphelion distance (2023 November 10). These
cases are much less likely for operational reasons, as the asteroid is
farther from Earth and thus the impact is more difficult to observe,
but are examined for completeness. The results in these cases
qualitatively resemble those of the 2022 September 1 impact, and are shown
in Figure~\ref{fig:otherdates} (Supplementary material). The 10 $\mu$m
particles arrive after thousands of years, larger particles may not
arrive at all, except in the case of the highest ejection speeds where
the MOIDs almost immediately reach low values.

The differences between the scenarios can be understood broadly in
terms of how close the DART impact occurs to the Didymos-Earth
MOID. This is because the ejecta particles, once released, travel on
(essentially) Keplerian ellipses and so necessarily return to the
point of impact, regardless of their ejection velocities.  So, though
the ejecta moves out on diverging heliocentric orbits after the
impact, the closed nature of these orbits brings them back to converge
on the impact point again. As a result, the meteoroid stream has its
smallest cross-section at the impact point.  And when the impact
occurs near the Didymos-Earth MOID, the resulting meteoroid stream is
therefore narrowest at its closest point of approach to Earth's orbit
(see e.g. Animated Figure~\ref{fig:animated}b).

Perhaps counter-intuitively, an impact {\it farther} from the
Earth-Didymos MOID can more easily produce a debris stream that has
{\it smaller} MOIDs with respect to Earth, and therefore takes a {\it
  shorter} time to be perturbed onto orbits that can reach our
planet. Since Earth reaches its MOID with Didymos in early
November\footnote{The close approach between Earth and Didymos on 2022
  October 4 mentioned in Section~\ref{sec:ejecta} occurs near but not
  exactly at the MOID.}, impact scenarios closer to this date produce
debris streams that take longer to reach Earth.

What does it take to get particles to Earth quickly at the nominal 1
Oct 2022 impact date? Additional simulations done with higher ejection
velocities show that all sizes will begin to arrive 15-30 days after
the DART impact, but only at ejection speeds of 6 km/s or more. These
are discussed in more detail later in Section~\ref{sec:direct}, but if
Didymos ejecta were observed in near-Earth space soon after impact
that would indicate that the ejection speeds are higher than expected
or that some other unmodeled process is at work.

What happens in the case of the 2022 September 1 impact where ejection
speeds reach 1000 m/s? When does this material arrive at Earth?  All
sizes except 10 $\mu$m arrive two years later when Didymos returns to
closest approach, and this will be detailed in Section~\ref{sec:2yearslater}.

\subsection{Fluxes}

In addition to whether or not material reaches Earth, there is the
question of how much.  The amount of material delivered is extremely
low, but for completeness we estimate the flux.  To do this in
detail would require a knowledge of the amount, size distribution and
velocity distribution of the ejecta, which are not well-known. Instead
we will adopt two simplistic models here, which will hopefully capture
the essential details without entangling them too much with
assumptions about the ejecta production process.

In our first 'nominal' model, we will assume that 1\% of the mass of
ejecta produced in the creation of a 10 m diameter crater is converted
entirely into particles of the size and ejection velocity being
discussed. This is certainly an over-estimate of the amount of
material expected when discussing larger, faster moving particles, and
an under-estimate for smaller and/or slower-moving particles.

For our second 'edge-case' model, we will assume that 1\% of all the
mass of Didymoon is converted entirely into particles of a particular
size and ejection velocity.  Such catastrophic scenario is extremely
unlikely, but we consider it here for two reasons. First, simulated
impacts have produced craters over 100~m in diameter on the 150~m
diameter moon \citep{stiatcbar15}, though for admittedly less probable
asteroid properties.  Secondly, the kinetic energy of the DART
impactor (assuming a 300 kg mass impacting at 6 km/s
\cite{micchekup16, cherivmic18}) exceeds the gravitational binding
energy of Didymoon as well as its orbital potential energy with its
primary by one to two orders of magnitude, so there is no strong
physical constraint against such an ejection event. Though unlikely,
the edge-case scenario is designed to capture the result of unusual
asteroid properties or other as-yet-unexplored eventualities
e.g. navigation malfunction resulting in an oblique impact; that might
create much more debris than expected. However even in this extreme
case, very little material ever reaches Earth.

For our nominal case, 1\% of the mass of a half-sphere of diameter 10~m is converted into
$N$ particles of diameter $d$, where
\begin{equation}
N_{nominal} \approx \frac{0.01}{2} \left( \frac{\rm{10~m}}{d} \right)^3 = 5 \times 10^{9} \left( \frac{d}{\rm{1~mm}} \right)^{-3} \label{eq:nominal}
\end{equation}
while for the edge-case scenario
\begin{equation}
N_{edge} \approx 0.01 \left( \frac{163~\rm{m}}{d} \right)^3 = 4.3 \times 10^{13} \left( \frac{d}{\rm{1~mm}} \right)^{-3} \label{eq:edgecase}
\end{equation}
where we have taken the secondary's diameter to be 163~m
\citep{micchekup16}. We will concentrate on a particle diameter of
1~mm because these are typical of sizes routinely detected by
meteor sensors on Earth and the low arrival speeds of DART-produced
meteors at Earth means that most meteor detection systems will have
difficulty detecting smaller particles. Of course, the simple nature
of our adopted size model means that numbers reported at millimeter sizes
can be scaled easily to any other desired ejecta model.

\subsubsection{Impact on 1 Oct 2022, direct arrival} \label{sec:direct}
Though an ejection speed of even 1000 m/s does not deliver ejecta
directly to Earth, additional simulations with faster ejecta
show that it can reach our planet at speeds of 6 km/s or higher. A
simulation of ejecta released at 5, 6, 7 and 8 km/s is shown in
Figure~\ref{fig:animated}a.  In this case debris arrives
directly from Didymos, and is roughly uniformly distributed
on an expanding sphere. As a result,  the flux $f$ is given approximately by
\begin{equation}
f \approx \frac{\alpha N}{4 \pi R^2 \Delta t}   \approx  3 \times 10^{-18} N  \left( \frac{\alpha}{2} \right) \left( \frac{R}{\rm{0.1~AU}} \right) ^{-2} \left( \frac{\Delta t}{{\rm 10~days}} \right)^{-1} {\rm km}^{-2} {\rm h}^{-1}\\
\end{equation}
where $N$ is the number of particles produced at the ejection velocity
in question, $R$ is the Earth-Didymos distance when the debris reaches
our planet ($\sim 0.1$~AU at 15-30 days after 2022 October 1) and
$\Delta t$ is the time the shell of ejecta takes to cross near-Earth
space, about 10 days. The parameter $\alpha$ is one over the fraction
of the sphere into which material is ejected. Our simulations assumed
uniform spherical ejection to allow for all ejection directions to be
assessed, but in practice the impact will release material only into a
half-sphere ($\alpha = 2$, adopted here), or perhaps an even narrower
cone ($\alpha > 2$). The nominal flux at Earth under these conditions
is $\sim 10^{-8}$~km$^{-2}$~h$^{-1}$, or for the edge-case $\sim
10^{-4}$~km$^{-2}$~h$^{-1}$.

For comparison, the background sporadic meteor flux at millimeter
sizes is $0.18 \pm 0.04$~km$^{-2}$~hr$^{-1}$ \citep{cambra11} as
measured by video meteor techniques, while a weak meteor shower might be
two orders of magnitude less e.g. the 2016 Camelopardalids
\citep{camblakin16}. So the DART-produced fluxes are certainly low:
could they be detected?

A typical meteor radar or camera has an effective collecting area
$\sim 1000$~km$^{2}$ \citep{werbro04}. So in the nominal case, such a
meteor detection system would see perhaps one meteor from Didymos during
the 10 days in question, and even in the edge case scenario it would
only see tens of meteors over that time span. Clearly, any detectable
meteor activity at Earth would indicate both exceptionally fast
and exceptionally abundant ejecta.

As the impact takes place below the ecliptic plane with Didymos rising
towards its ascending node, any directly-delivered meteors arrive from
southerly radiant directions. The high ejection speeds, together with
their low velocities relative to Earth (4-12 km/s, peaking at 9),
result in a very broad radiant extending for tens of degrees centered
roughly at RA=0$\degree$, Dec = -50$\degree$. These meteors will be
fainter than usual for their size due to the rapid drop in ionization
production \citep{jon97,delmuntho18} and light production
\citep{subcam18} by meteors at such low speeds.

The calculations above  ignore gravitational focusing, which increases the
effective cross-section of Earth by a factor of $1+(v_{esc}/v)^2$,
where $v$ is the meteoroid arrival speed, and $v_{esc}$ is the escape
speed of Earth ($\approx 8$~km/s). Though the low arrival speed means
that gravitational focusing is not completely negligible, it only
increases the fluxes determined above by a factor $\approx 2$.

\subsubsection{Impact on 2022 September 1, arrival two years later} \label{sec:2yearslater}

Next we consider the flux of 1000 m/s ejecta released during an impact
on 2022 September 1. This material disperses quickly along the orbit and
arrives at Earth at about the same time as Didymos' close approach in
2024, as can be seen in Fig.~\ref{fig:animated}b.

To calculate the flux we will assume that the particles have dispersed
uniformly around their orbit (as is evident from
Fig.~\ref{fig:animated}b), and occupy a toroidal volume $V$ around
their mean orbit such that $V \approx 2 \pi a A$, where $a$ is
Didymos' semimajor axis ($\approx 1.64$~AU), and $A$ is the stream
cross-sectional area. The area $A$ is computed from an ellipse fitted to
a cross-section of the simulated stream at its closest approach to
Earth's orbit. Then the particle number density $n$ is just $N/V$ and
the flux is $n v_{rel}$ where $v_{rel}$ is the relative velocity of
the meteoroids to Earth, $\approx 5$~km/s.

For all particle sizes at 1000 m/s ejection speed, the
cross-section is $\approx 3 \times 10^{-3}$~AU$^2$, and the nominal
flux at Earth $\sim 10^{-9}$~km$^{-2}$~h$^{-1}$. This is an order of
magnitude below that determined for direct arrival in
section~\ref{sec:direct}, and even for our edge case scenario, a
detector with an effective collecting area of $1000$~km$^2$ would see
$\las$~1 millimeter-sized meteor per day. As a result, it is not
expected that this extremely weak meteor activity will be detected.

For completeness we note that any meteoroids that do reach Earth will
be accelerated by Earth's gravity to an in-atmosphere speed of 12
km/s, and would arrive from a broad geocentric radiant in the vicinity
of RA = $295\degree$ Dec= $-40\degree$. These meteors will also be
fainter than typical for their sizes both in radar and optical meteor
systems due to their low speeds (see Section~\ref{sec:direct}).

\subsubsection{Flux in the new meteoroid stream}

\begin{figure}
\plotone{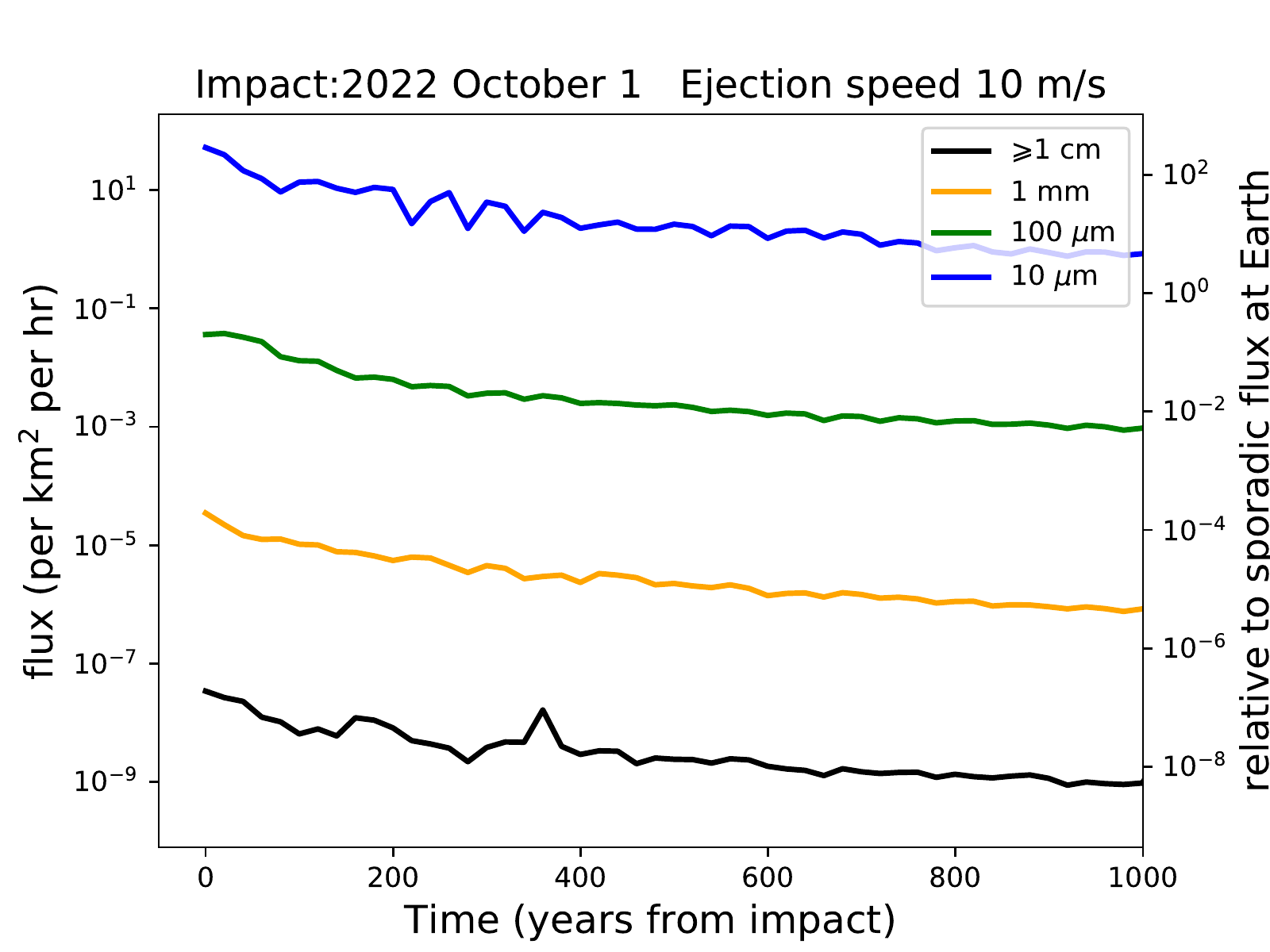}
\caption{Estimated flux near Didymos' orbit at its MOID with Earth, for the nominal ejected
  mass at a 10 m/s ejection speed. The right-hand y-axis gives the
  flux relative to the sporadic meteoroid flux at Earth of 0.18 per
  km$^2$ per hour \citep{cambra11}. \label{fig:awayfromEarth}}
\end{figure}

Most of the debris from a DART impact will not arrive at Earth but
will disperse into a meteoroid stream near Didymos' orbit. What flux
can a spacecraft flying through this stream expect? We estimate this
by the same technique used in section~\ref{sec:2yearslater}. For
consistency we will use the stream cross-section determined from the
simulations at the Earth MOID, and the same relative velocity (5~km/s)
considered earlier, but a more correct determination would require
considering the specific speed and position of the craft as it crosses
the stream.  The predicted fluxes for the nominal case assuming a 10
m/s ejection speed, which produces the highest flux values, are shown
in Figure~\ref{fig:awayfromEarth}.

The figure shows that at millimeter sizes, the nominal flux will be
$\sim 10^{-5}$~km$^{-2}$~h$^{-1}$ initially, and dropping over time as
the stream evolves. This is still quite low, though orders of
magnitude higher than the fluxes discussed earlier. The increased flux
is a result of a smaller cross-section, resulting from the lower
ejection speed.  The initial cross-section at the Earth MOID for 10
m/s ejection speeds is $\las 10^{-7}$~AU$^2$ for all particle sizes,
about four orders of magnitude smaller than in the 1000 m/s ejection
case, and producing a 10$^{4}$ increase in the meteoroid flux. This
demonstrates that low-speed ejecta, which will be far more abundant,
may be of more long-term concern. It is also relevant to the design of
asteroid mining operations which may inadvertently or deliberately
release debris at low speeds, and which could create dense meteoroid
streams in their vicinity.

In our edge-case scenario, the initial flux of millimeter-sized
material in the stream would actually exceed that associated with weak
meteor showers by a factor of 100, reaching levels comparable to the
background sporadic meteoroid flux at Earth (this scenario corresponds
to the 1 mm line in Figure~\ref{fig:awayfromEarth}, multiplied by
$N_{edge}/N_{nominal} \approx 10^4$).  Though the risk to spacecraft
even in this case remains low, it is conceivable that DART or perhaps
more likely, future more ambitious planetary defense tests, will
result in the production of meteoroid streams where the debris fluxes
exceed those naturally occurring within the Solar System. These
streams carry implications for the safety of spacecraft that need to
cross them, and though associated risks are likely to be initially
very low, they will undoubtedly increase with time much as has the
orbital debris problem in low Earth orbit.

We note that the flux values of Figure~\ref{fig:awayfromEarth} assume
the particles' are fully dispersed around the stream's mean orbit. This
dispersion takes longer in the case of low ejection speeds ($\approx
250$~yr for all particle sizes at 10~m/s ejection speed).  As a
result, the flux will initially be higher along some portions of Didymos' orbit
and lower in others. Determining the actual debris flux encountered by
a spacecraft crossing the stream in the near-future, such as the Hera
spacecraft planned to observe the effects of the DART impact, would
require a more detailed study than is done here.

\section{Conclusions}

Debris ejected by the DART impact on Didymoon may reach Earth
in small numbers. Ejecta can reach Earth directly within 15-30 days
after impact if the ejection speeds reach 6~km/s, though these speeds
are higher than expected. The debris cloud will subsequently spread
out into a meteoroid stream.  The baseline DART impact date of 1
October 2022 does not produce a stream that crosses Earth's orbit,
at least not immediately, though its dynamical evolution will
eventually bring some of the debris to near-Earth space after
thousands of years.

Other impact dates can place material onto orbits which immediately
cross that of Earth, though only at high (1000 m/s) ejection
speeds, and only a very small amount of the ejecta is expected to
reach our planet.

The meteoroid stream produced by the impact remains primarily in the
vicinity of Didymos' orbit. The stream's cross-section is larger for
larger ejection speeds; as a result, low-speed ejecta, expected to be
relatively abundant, produce a denser meteoroid stream.  Though it is
unlikely to occur in the case of the DART impact, future human
asteroid operations such as planetary defense tests or asteroid
mining, could conceivably produce debris streams whose meteoroid
particle content rivals or exceeds naturally-occurring meteoroid
streams.  Streams initially emplaced far from Earth may reach
near-Earth space after hundreds or thousands of years, and thus
require some long-term planning. Though such a stream would have to be
quite dense and contain a large number of decameter or larger class
asteroids to be dangerous to the Earth's surface, a much lower density
of small particles could be inconvenient or detrimental to some space
operations.  JWST has a large vulnerable mirror and future space
telescopes are likely to be even more ambitious and sensitive. The
Gaia spacecraft attitude control system already has to deal with
natural meteoroid impacts \citep{sermilmar16}, as does LISA
Pathfinder's \citep{thoslubak19}. Though one is tempted to dismiss the
problem as negligible at this time, it is reminiscent of the problem
of space debris in low Earth orbit. Neglected initially, we are now
reaching a point where we may be denied the full use of valuable
portions of near-Earth space because of orbital debris build-up. Much
future expense and risk could be averted if the same story does not
unfold with asteroidal debris production.

\acknowledgments

The author thanks the reviewers for thoughtful comments that much
improved this manuscript.  Funding for this work was provided through NASA
co-operative agreement 80NSSC18M0046 and the Natural Sciences and
Engineering Research Council (NSERC) of Canada (Grant
no. RGPIN-2018-05659).

%





\newpage
\appendix

\section{Supplementary information}

\begin{figure}[hb]
\gridline{\fig{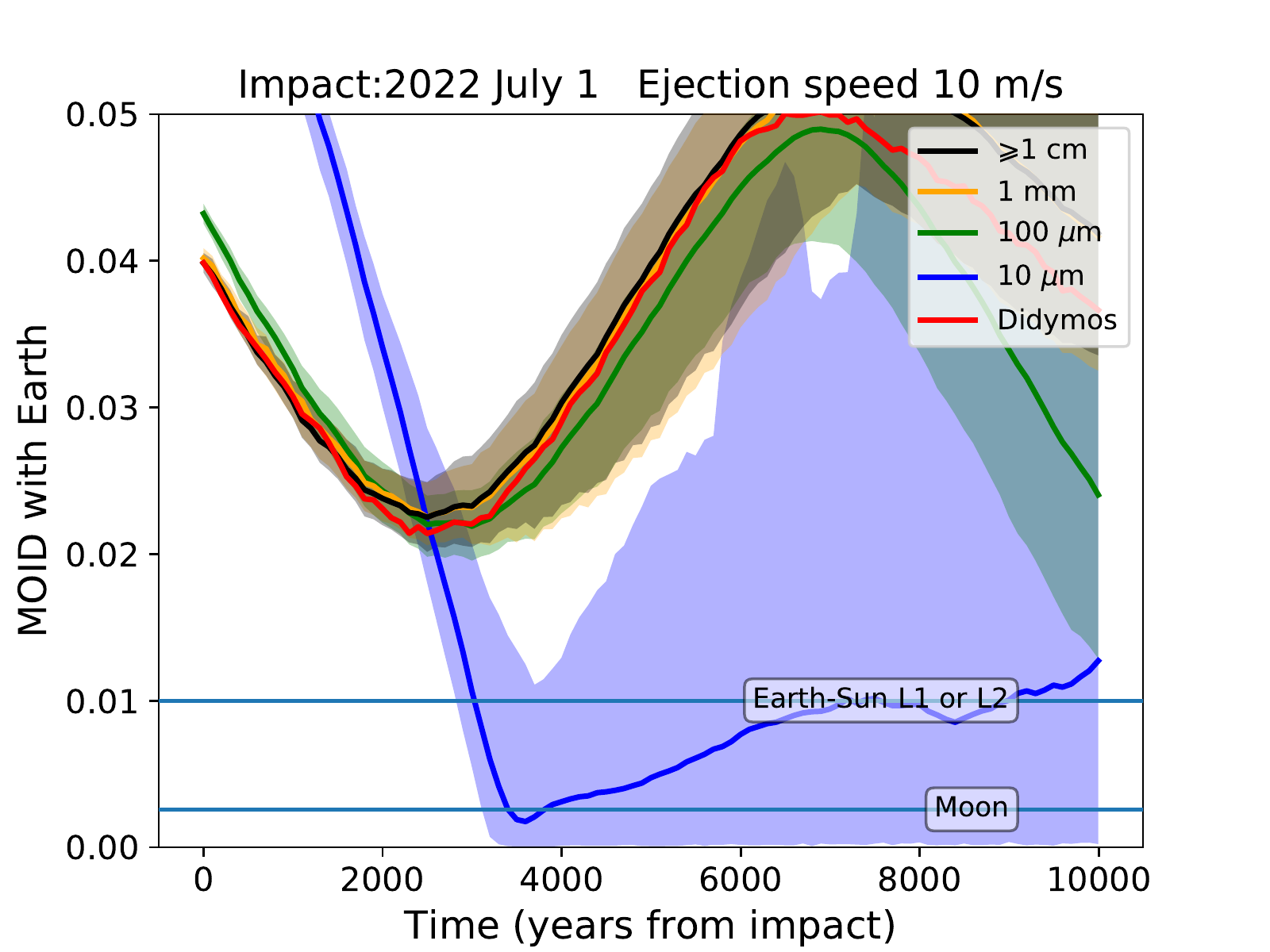}{0.33\textwidth}{(a)}
          \fig{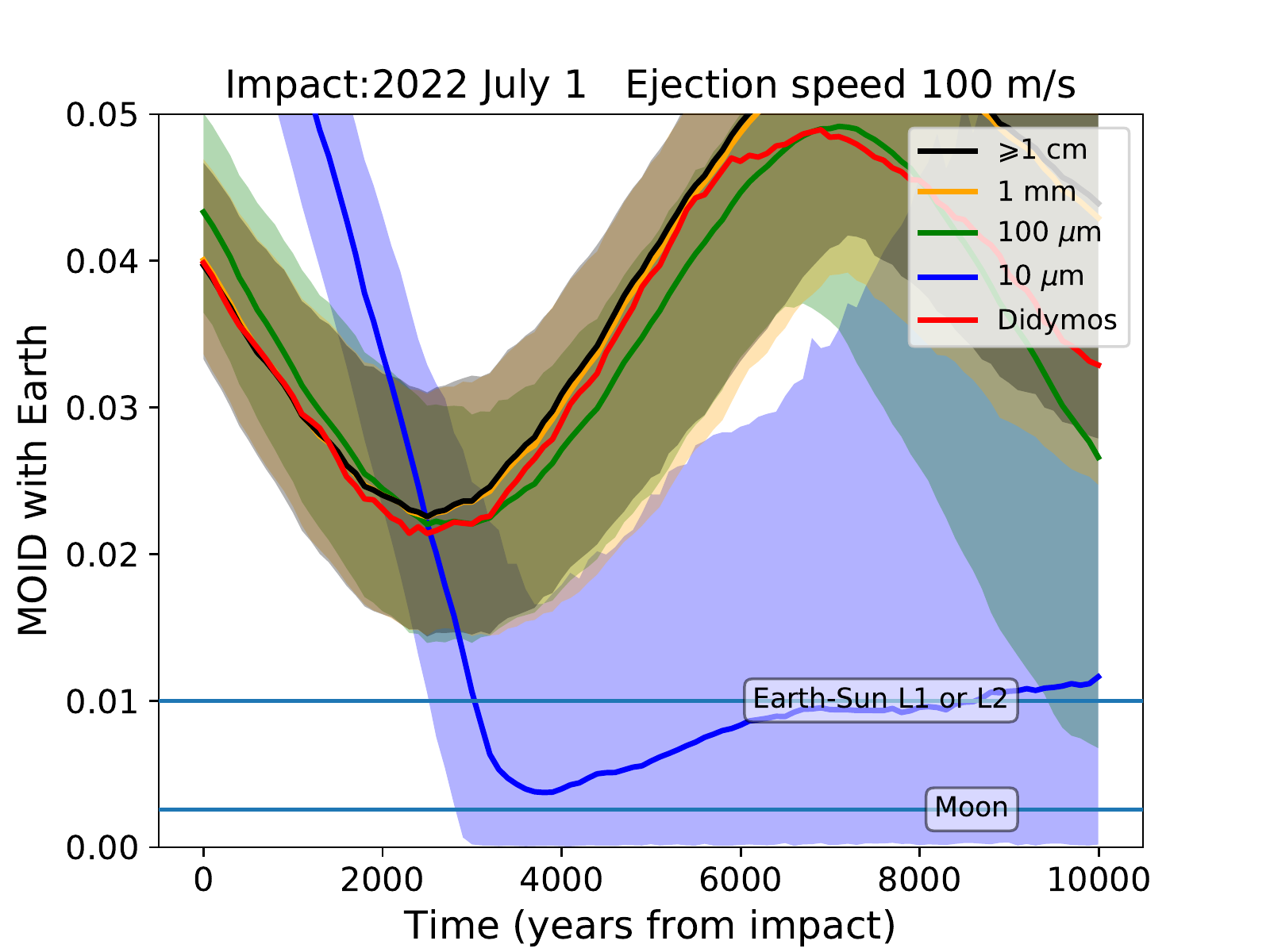}{0.33\textwidth}{(b)}
          \fig{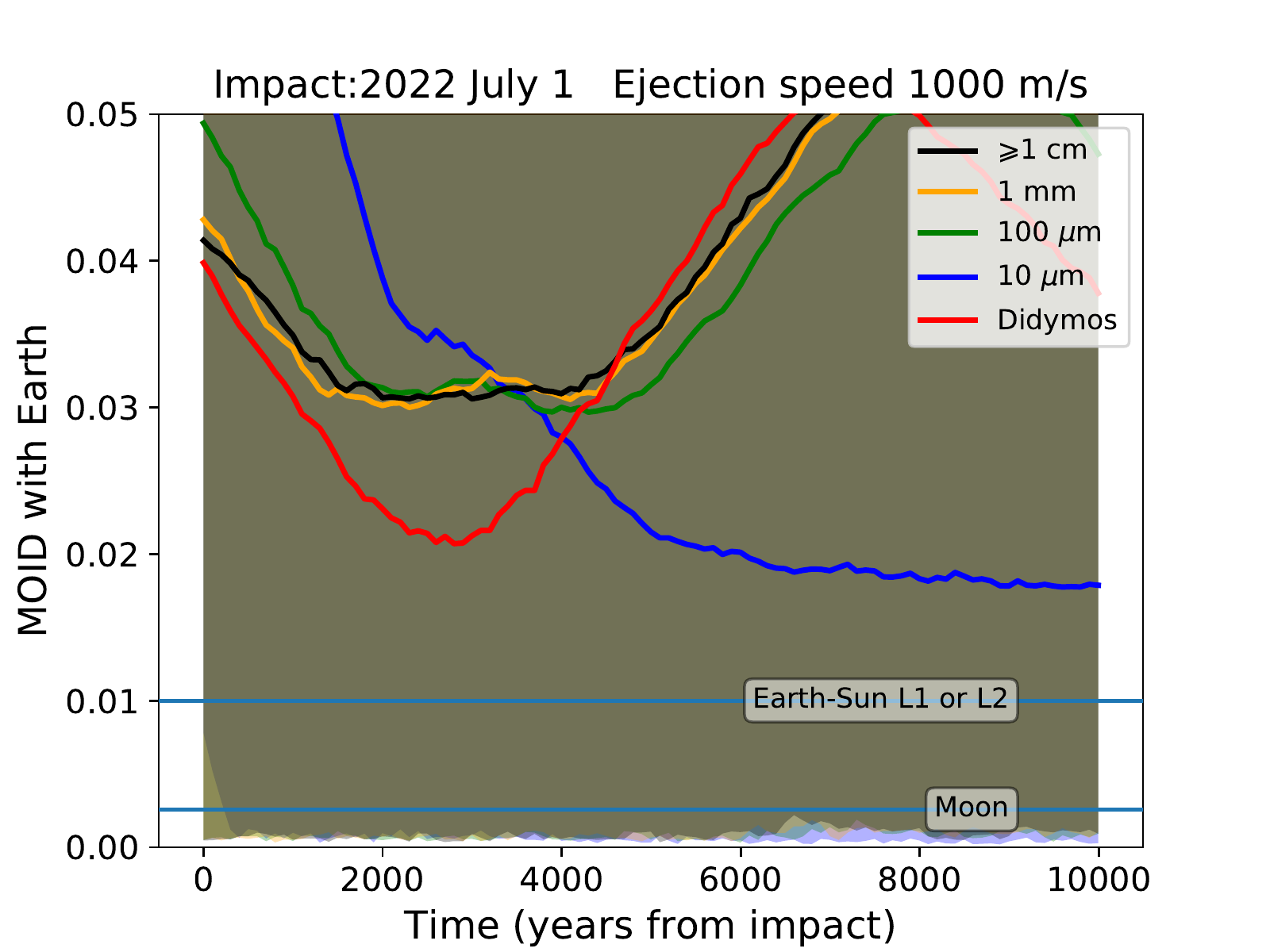}{0.33\textwidth}{(c)}
          }
\gridline{\fig{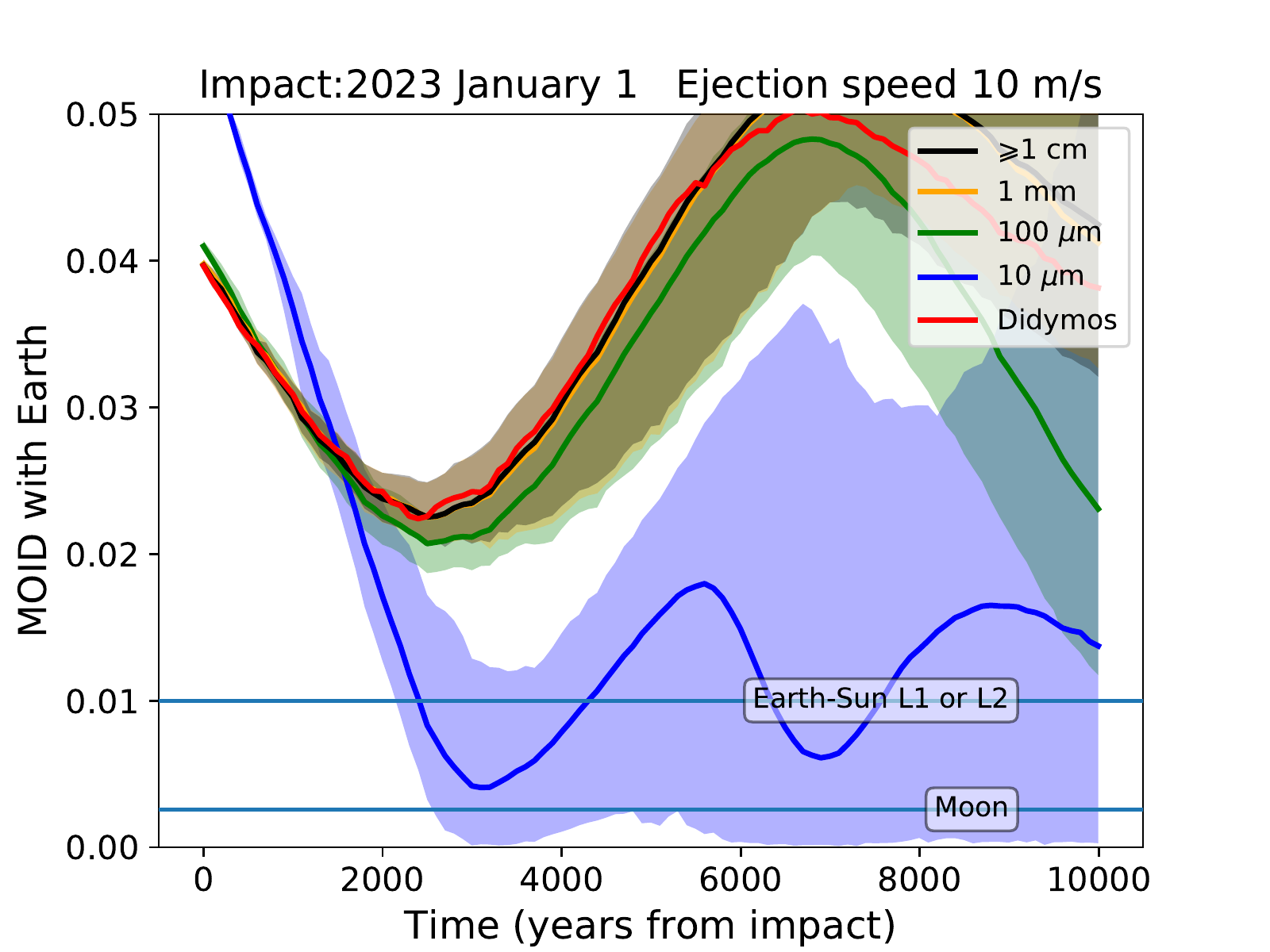}{0.33\textwidth}{(a)}
          \fig{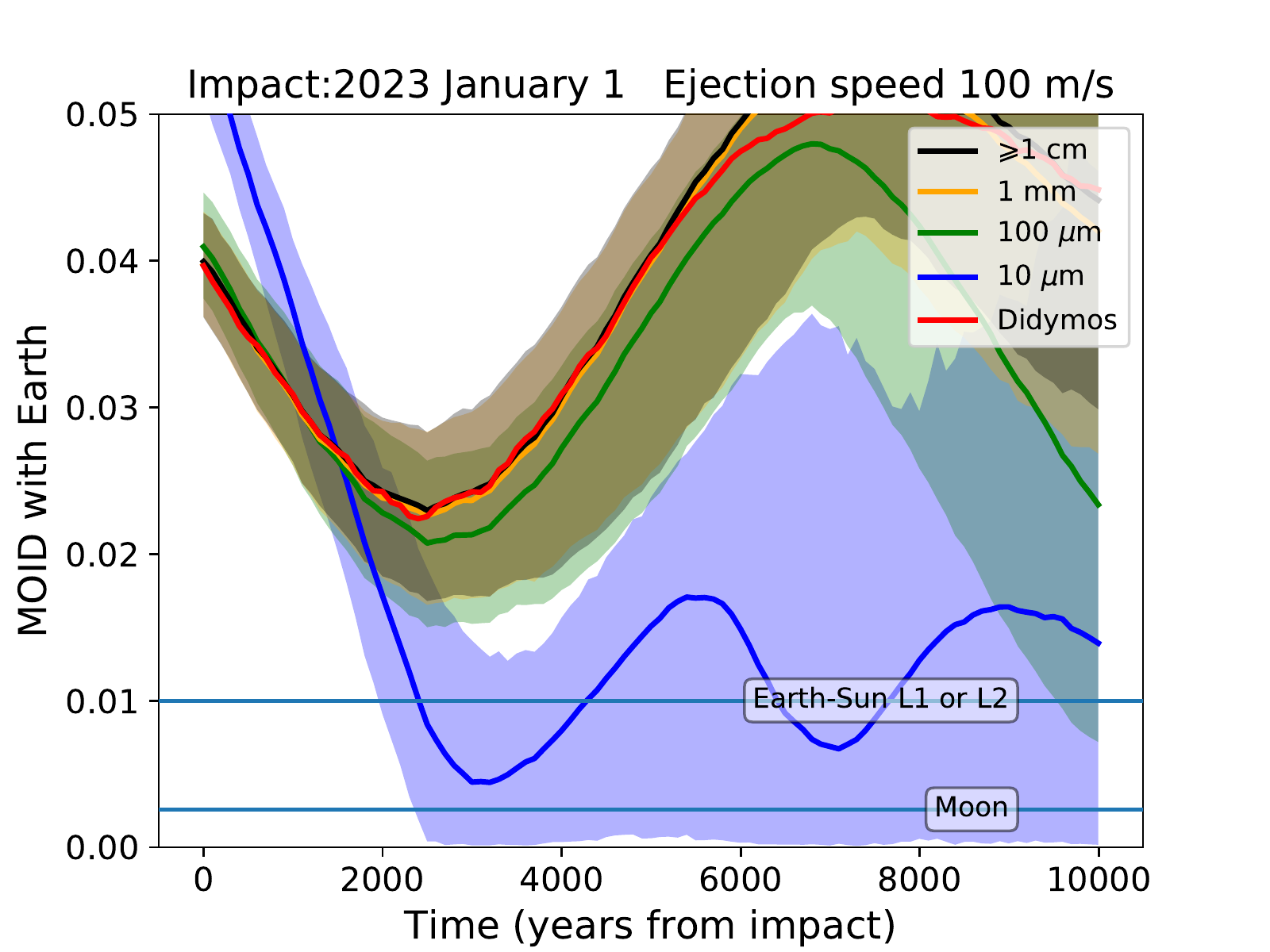}{0.33\textwidth}{(b)}
          \fig{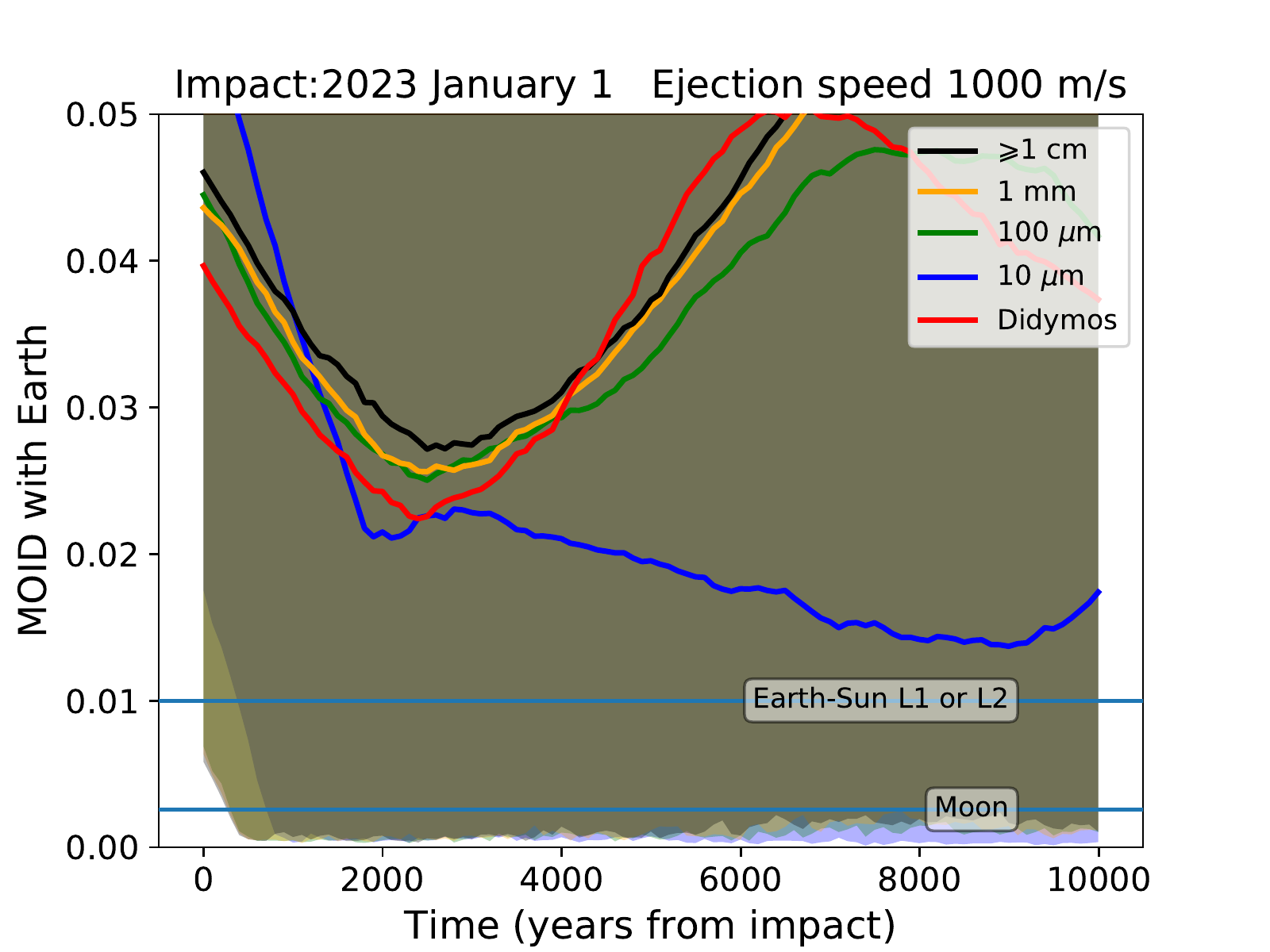}{0.33\textwidth}{(c)}
          }
\gridline{\fig{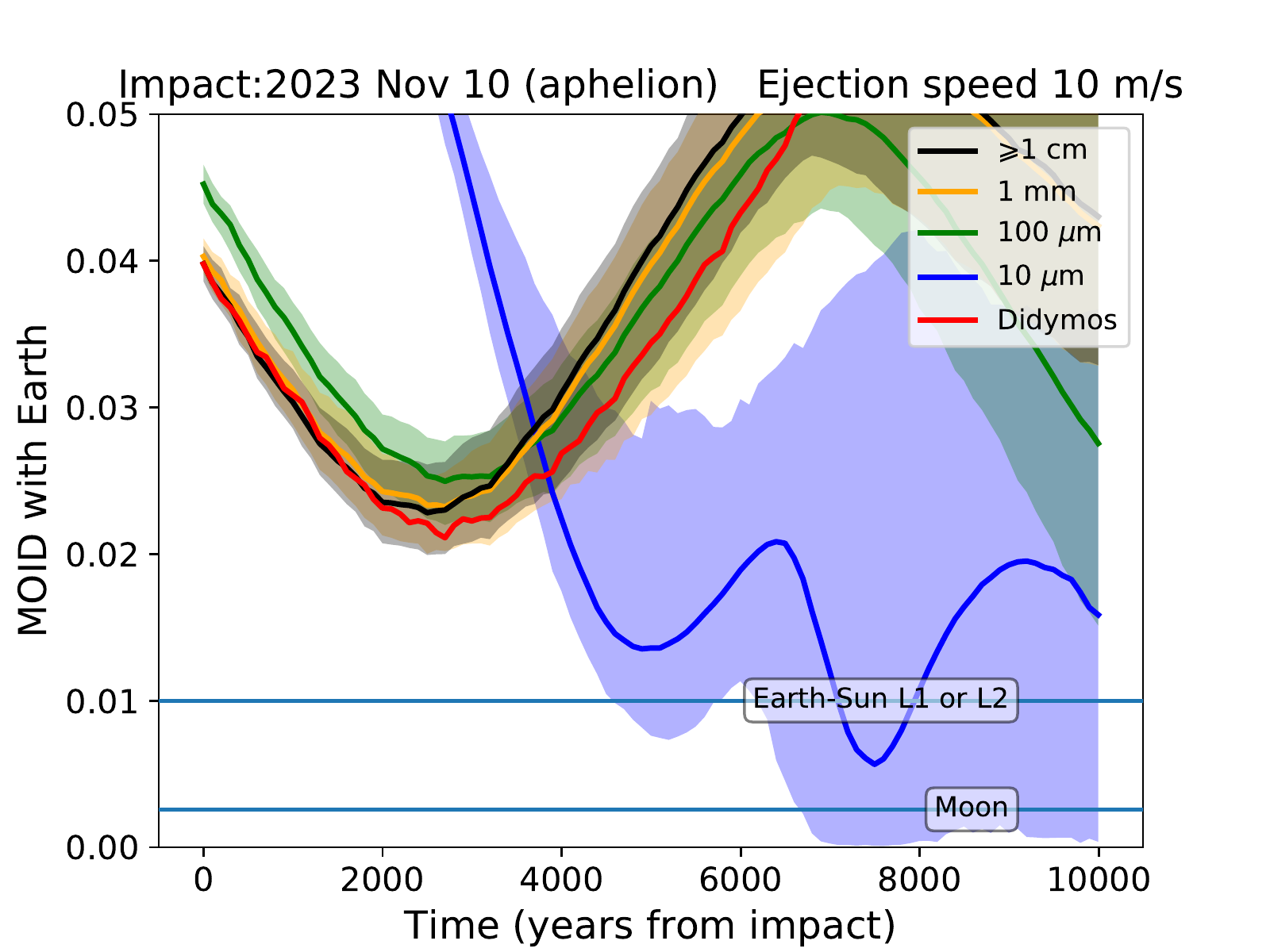}{0.33\textwidth}{(a)}
          \fig{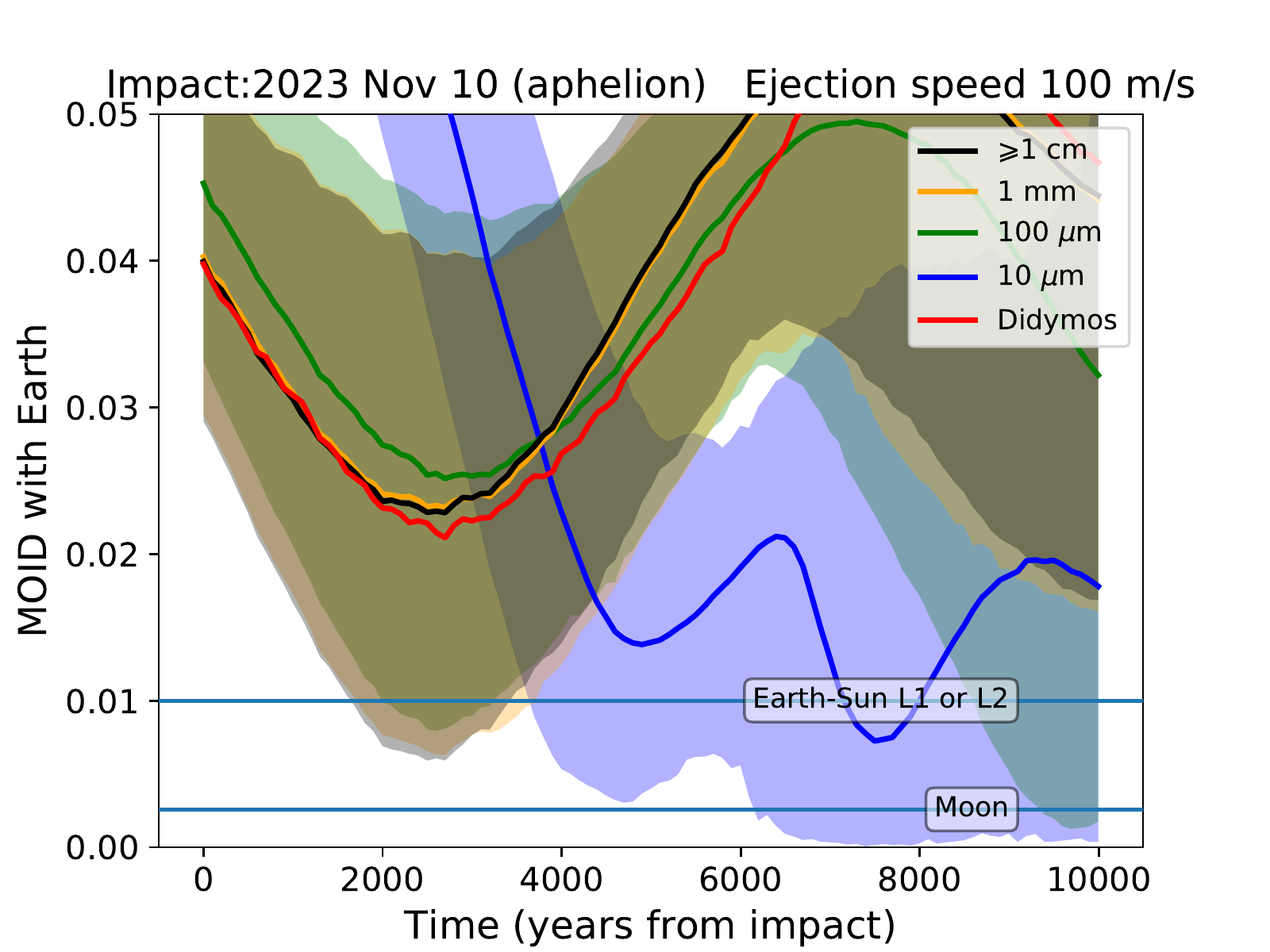}{0.33\textwidth}{(b)}
          \fig{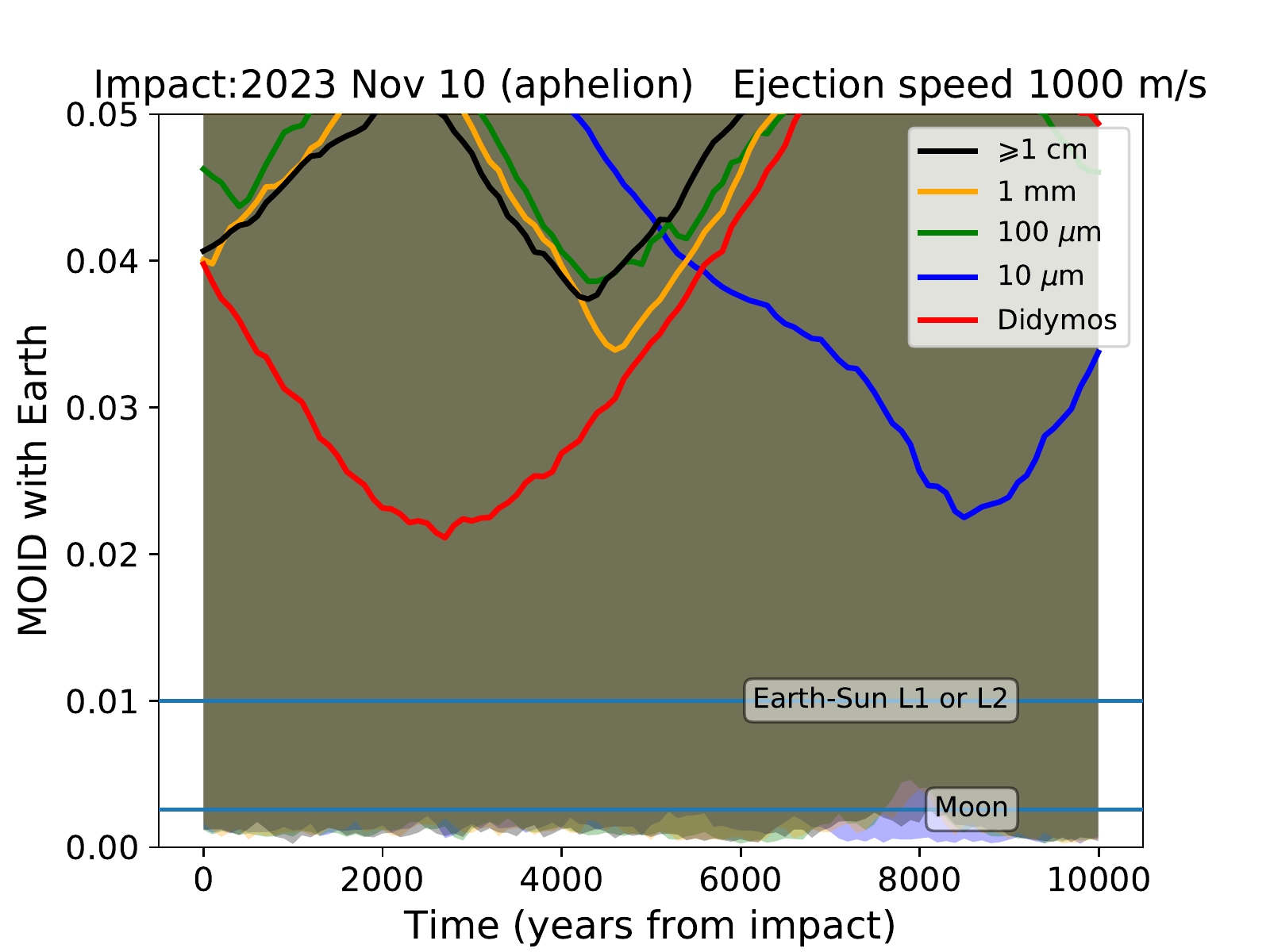}{0.33\textwidth}{(c)}
          }
\caption{Supplementary figure. The evolution of the MOIDs of the simulated DART-ejected debris in the case of an impact on dates 3 months on either side of the nominal date of 2022 October 1 (2022 July 1 and 2023 January 1), as well as at Didymos' subsequent aphelion (2023 November 10).  See Figure~\ref{fig:1Oct2022} and the text for more details.\label{fig:otherdates}}
\end{figure}



\bibliographystyle{aasjournal}

\bibliography{Wiegert}

\begin{thebibliography}{}
\expandafter\ifx\csname natexlab\endcsname\relax\def\natexlab#1{#1}\fi
\providecommand{\url}[1]{\href{#1}{#1}}
\providecommand{\dodoi}[1]{doi:~\href{http://doi.org/#1}{\nolinkurl{#1}}}
\providecommand{\doeprint}[1]{\href{http://ascl.net/#1}{\nolinkurl{http://ascl.net/#1}}}
\providecommand{\doarXiv}[1]{\href{https://arxiv.org/abs/#1}{\nolinkurl{https://arxiv.org/abs/#1}}}

\bibitem[{{A'Hearn} {et~al.}(2005){A'Hearn}, {Belton}, {Delamere}, {Kissel},
  {Klaasen}, {McFadden}, {Meech}, {Melosh}, {Schultz}, {Sunshine}, {Thomas},
  {Veverka}, {Yeomans}, {Baca}, {Busko}, {Crockett}, {Collins}, {Desnoyer},
  {Eberhardy}, {Ernst}, {Farnham}, {Feaga}, {Groussin}, {Hampton}, {Ipatov},
  {Li}, {Lindler}, {Lisse}, {Mastrodemos}, {Owen}, {Richardson}, {Wellnitz}, \&
  {White}}]{ahebeldel05}
{A'Hearn}, M.~F., {Belton}, M.~J.~S., {Delamere}, W.~A., {et~al.} 2005,
  Science, 310, 258, \dodoi{10.1126/science.1118923}

\bibitem[{Atkinson {et~al.}(2000)Atkinson, Tickell, \& Williams}]{atkticwil00}
Atkinson, H., Tickell, C., \& Williams, D. 2000, Report of the Task Force on
  potentially hazardous Near Earth Objects

\bibitem[{{Binzel}(2000)}]{bin00}
{Binzel}, R.~P. 2000, \planss, 48, 297, \dodoi{10.1016/S0032-0633(00)00006-4}

\bibitem[{{Blaauw} {et~al.}(2011){Blaauw}, {Campbell-Brown}, \&
  {Weryk}}]{blacamwer11}
{Blaauw}, R.~C., {Campbell-Brown}, M.~D., \& {Weryk}, R.~J. 2011, MNRAS, 412,
  2033, \dodoi{10.1111/j.1365-2966.2010.18038.x}

\bibitem[{{Britt} {et~al.}(2002){Britt}, {Yeomans}, {Housen}, \&
  {Consolmagno}}]{briyeohou02}
{Britt}, D.~T., {Yeomans}, D., {Housen}, K., \& {Consolmagno}, G. 2002,
  Asteroids III, 485

\bibitem[{{Campbell-Brown} {et~al.}(2016){Campbell-Brown}, {Blaauw}, \&
  {Kingery}}]{camblakin16}
{Campbell-Brown}, M.~D., {Blaauw}, R., \& {Kingery}, A. 2016, \icarus, 277,
  141, \dodoi{10.1016/j.icarus.2016.05.001}

\bibitem[{{Campbell-Brown} \& {Braid}(2011)}]{cambra11}
{Campbell-Brown}, M.~D., \& {Braid}, D. 2011, in Meteoroids: The Smallest Solar
  System Bodies, ed. W.~J. {Cooke}, D.~E. {Moser}, B.~F. {Hardin}, \&
  D.~{Janches}, 304--312

\bibitem[{{Chambers}(1999)}]{cha99}
{Chambers}, J.~E. 1999, MNRAS, 304, 793

\bibitem[{{Cheng} {et~al.}(2018){Cheng}, {Rivkin}, {Michel}, {Atchison},
  {Barnouin}, {Benner}, {Chabot}, {Ernst}, {Fahnestock}, {Kueppers}, {Pravec},
  {Rainey}, {Richardson}, {Stickle}, \& {Thomas}}]{cherivmic18}
{Cheng}, A.~F., {Rivkin}, A.~S., {Michel}, P., {et~al.} 2018, \planss, 157,
  104, \dodoi{10.1016/j.pss.2018.02.015}

\bibitem[{{Crifo}(1995)}]{cri95}
{Crifo}, J.~F. 1995, ApJ, 445, 470, \dodoi{10.1086/175712}

\bibitem[{{DeLuca} {et~al.}(2018){DeLuca}, {Munsat}, {Thomas}, \&
  {Sternovsky}}]{delmuntho18}
{DeLuca}, M., {Munsat}, T., {Thomas}, E., \& {Sternovsky}, Z. 2018, \planss,
  156, 111, \dodoi{10.1016/j.pss.2017.11.003}

\bibitem[{{Domingo} {et~al.}(1995){Domingo}, {Fleck}, \&
  {Poland}}]{domflepol95}
{Domingo}, V., {Fleck}, B., \& {Poland}, A.~I. 1995, \ssr, 72, 81,
  \dodoi{10.1007/BF00768758}

\bibitem[{{Everhart}(1985)}]{eve85}
{Everhart}, E. 1985, in Dynamics of Comets: Their Origin and Evolution, ed.
  A.~{Carusi} \& G.~B. {Valsecchi} (Dordrecht: Kluwer), 185--202

\bibitem[{{Fladeland} {et~al.}(2019){Fladeland}, {Boley}, \&
  {Byers}}]{flabolbye19}
{Fladeland}, L., {Boley}, A.~C., \& {Byers}, M. 2019, arXiv e-prints,
  arXiv:1911.12840.
\newblock \doarXiv{1911.12840}

\bibitem[{{Fujiwara} {et~al.}(2006){Fujiwara}, {Kawaguchi}, {Yeomans}, {Abe},
  {Mukai}, {Okada}, {Saito}, {Yano}, {Yoshikawa}, {Scheeres}, {Barnouin-Jha},
  {Cheng}, {Demura}, {Gaskell}, {Hirata}, {Ikeda}, {Kominato}, {Miyamoto},
  {Nakamura}, {Nakamura}, {Sasaki}, \& {Uesugi}}]{fujkawyeo06}
{Fujiwara}, A., {Kawaguchi}, J., {Yeomans}, D.~K., {et~al.} 2006, Science, 312,
  1330, \dodoi{10.1126/science.1125841}

\bibitem[{{Gardner} {et~al.}(2006){Gardner}, {Mather}, {Clampin}, {Doyon},
  {Greenhouse}, {Hammel}, {Hutchings}, {Jakobsen}, {Lilly}, {Long}, {Lunine},
  {McCaughrean}, {Mountain}, {Nella}, {Rieke}, {Rieke}, {Rix}, {Smith},
  {Sonneborn}, {Stiavelli}, {Stockman}, {Windhorst}, \& {Wright}}]{garmatcla06}
{Gardner}, J.~P., {Mather}, J.~C., {Clampin}, M., {et~al.} 2006, \ssr, 123,
  485, \dodoi{10.1007/s11214-006-8315-7}

\bibitem[{{Holsapple} \& {Housen}(2007)}]{holhou07}
{Holsapple}, K.~A., \& {Housen}, K.~R. 2007, \icarus, 187, 345,
  \dodoi{10.1016/j.icarus.2006.08.029}

\bibitem[{{Jones}(1995)}]{jon95}
{Jones}, J. 1995, MNRAS, 275, 773

\bibitem[{{Jones} {et~al.}(2005){Jones}, {Brown}, {Ellis}, {Webster},
  {Campbell-Brown}, {Krzemenski}, \& {Weryk}}]{jonbroell05}
{Jones}, J., {Brown}, P., {Ellis}, K.~J., {et~al.} 2005, Plan. Space Sci., 53,
  413

\bibitem[{{Jones}(1997)}]{jon97}
{Jones}, W. 1997, \mnras, 288, 995, \dodoi{10.1093/mnras/288.4.995}

\bibitem[{{Lauretta} {et~al.}(2019){Lauretta}, {Dellagiustina}, {Bennett},
  {Golish}, {Becker}, {Balram-Knutson}, {Barnouin}, {Becker}, {Bottke},
  {Boynton}, {Campins}, {Clark}, {Connolly}, {Drouet D'Aubigny}, {Dworkin},
  {Emery}, {Enos}, {Hamilton}, {Hergenrother}, {Howell}, {Izawa}, {Kaplan},
  {Nolan}, {Rizk}, {Roper}, {Scheeres}, {Smith}, {Walsh}, {Wolner}, \&
  {Osiris-Rex Team}}]{laudelben19}
{Lauretta}, D.~S., {Dellagiustina}, D.~N., {Bennett}, C.~A., {et~al.} 2019,
  \nat, 568, 55, \dodoi{10.1038/s41586-019-1033-6}

\bibitem[{{Michel} \& {Yu}(2017)}]{micyu17}
{Michel}, P., \& {Yu}, Y. 2017, in European Planetary Science Congress,
  EPSC2017--82

\bibitem[{{Michel} {et~al.}(2016){Michel}, {Cheng}, {K{\"u}ppers}, {Pravec},
  {Blum}, {Delbo}, {Green}, {Rosenblatt}, {Tsiganis}, {Vincent}, {Biele},
  {Ciarletti}, {H{\'e}rique}, {Ulamec}, {Carnelli}, {Galvez}, {Benner},
  {Naidu}, {Barnouin}, {Richardson}, {Rivkin}, {Scheirich}, {Moskovitz},
  {Thirouin}, {Schwartz}, {Campo Bagatin}, \& {Yu}}]{micchekup16}
{Michel}, P., {Cheng}, A., {K{\"u}ppers}, M., {et~al.} 2016, Advances in Space
  Research, 57, 2529, \dodoi{10.1016/j.asr.2016.03.031}

\bibitem[{{Moorhead} {et~al.}(2015){Moorhead}, {Koehler}, \&
  {Cooke}}]{mookoecoo15}
{Moorhead}, A.~V., {Koehler}, H.~M., \& {Cooke}, W.~J. 2015, NASA Meteoroid
  Engineering Model Release 2.0, Tech. rep., NASA

\bibitem[{{Morrison} {et~al.}(2004){Morrison}, {Chapman}, {Steel}, \&
  {Binzel}}]{morchaste04}
{Morrison}, D., {Chapman}, C.~R., {Steel}, D., \& {Binzel}, R.~P. 2004, in
  Mitigation of Hazardous Comets and Asteroids, ed. M.~J.~S. {Belton}, T.~H.
  {Morgan}, N.~H. {Samarasinha}, \& D.~K. {Yeomans}, 353

\bibitem[{Raducan {et~al.}(2019)Raducan, Davison, \& Collins}]{raddavcol19}
Raducan, S., Davison, T., \& Collins, G. 2019, Planetary and Space Science,
  104756, \dodoi{https://doi.org/10.1016/j.pss.2019.104756}

\bibitem[{{Richardson} \& {O'Brien}(2016)}]{ricobr16}
{Richardson}, J.~E., \& {O'Brien}, D.~P. 2016, in AAS/Division for Planetary
  Sciences Meeting Abstracts \#48, AAS/Division for Planetary Sciences Meeting
  Abstracts, 329.06

\bibitem[{{Serpell} {et~al.}(2016){Serpell}, {Milligan}, {Marie}, \&
  {Collins}}]{sermilmar16}
{Serpell}, E., {Milligan}, D., {Marie}, J., \& {Collins}, P. 2016, in ,
  Meteoroids 2016 Meeting Abstracts

\bibitem[{{Stickle} {et~al.}(2015){Stickle}, {Atchison}, {Barnouin}, {Cheng},
  {Ernst}, {Richardson}, \& {Rivkin}}]{stiatcbar15}
{Stickle}, A.~M., {Atchison}, J.~A., {Barnouin}, O.~S., {et~al.} 2015, in
  AAS/Division for Planetary Sciences Meeting Abstracts \#47, AAS/Division for
  Planetary Sciences Meeting Abstracts, 312.14

\bibitem[{Stickle {et~al.}(2020)Stickle, Syal, Cheng, Collins, Davison, Gisler,
  G{\"u}ldemeister, Heberling, Luther, Michel, {et~al.}}]{stisyache20}
Stickle, A.~M., Syal, M.~B., Cheng, A.~F., {et~al.} 2020, Icarus, 338, 113446

\bibitem[{Stokes {et~al.}(2003)Stokes, Yeomans, Bottke, Chesley, Evans, Gold,
  Harris, Jewitt, Kelso, McMillan, Spahr, \& Worden}]{stoyeobot03}
Stokes, G., Yeomans, D., Bottke, W., {et~al.} 2003, Study to Determine the
  Feasibility of Extending the Search for Near-Earth Objects to Smaller
  Limiting Diameters

\bibitem[{{Subasinghe} \& {Campbell-Brown}(2018)}]{subcam18}
{Subasinghe}, D., \& {Campbell-Brown}, M. 2018, \aj, 155, 88,
  \dodoi{10.3847/1538-3881/aaa3e0}

\bibitem[{{Tauber} {et~al.}(2004){Tauber}, {ESA Scientific Collaboration}, \&
  {Planck Scientific Collaboration}}]{tau04}
{Tauber}, J.~A., {ESA Scientific Collaboration}, \& {Planck Scientific
  Collaboration}. 2004, Advances in Space Research, 34, 491,
  \dodoi{10.1016/j.asr.2003.05.025}

\bibitem[{{Thorpe} {et~al.}(2019){Thorpe}, {Slutsky}, {Baker}, {Littenberg},
  {Hourihane}, {Pagane}, {Pokorny}, {Janches}, {LISA Pathfinder Collaboration},
  {Armano}, {Audley}, {Auger}, {Baird}, {Bassan}, {Binetruy}, {Born},
  {Bortoluzzi}, {Brandt}, {Caleno}, {Cavalleri}, {Cesarini}, {Cruise},
  {Danzmann}, {de Deus Silva}, {De Rosa}, {Di Fiore}, {Diepholz}, {Dixon},
  {Dolesi}, {Dunbar}, {Ferraioli}, {Ferroni}, {Fitzsimons}, {Flatscher},
  {Freschi}, {Garc{\'\i}a Marirrodriga}, {Gerndt}, {Gesa}, {Gibert},
  {Giardini}, {Giusteri}, {Grado}, {Grimani}, {Grzymisch}, {Harrison},
  {Heinzel}, {Hewitson}, {Hollington}, {Hoyland}, {Hueller}, {Inchausp{\'e}},
  {Jennrich}, {Jetzer}, {Johlander}, {Karnesis}, {Kaune}, {Korsakova},
  {Killow}, {Lobo}, {Lloro}, {Liu}, {L{\'o}pez-Zaragoza}, {Maarschalkerweerd},
  {Mance}, {Mart{\'\i}n}, {Martin-Polo}, {Martino}, {Martin-Porqueras},
  {Madden}, {Mateos}, {McNamara}, {Mendes}, {Mendes}, {Nofrarias},
  {Paczkowski}, {Perreur-Lloyd}, {Petiteau}, {Pivato}, {Plagnol}, {Prat},
  {Ragnit}, {Ramos-Castro}, {Reiche}, {Robertson}, {Rozemeijer}, {Rivas},
  {Russano}, {Sarra}, {Schleicher}, {Shaul}, {Sopuerta}, {Stanga}, {Sumner},
  {Texier}, {Trenkel}, {Tr{\"o}bs}, {Vetrugno}, {Vitale}, {Wanner}, {Ward},
  {Wass}, {Wealthy}, {Weber}, {Wissel}, {Wittchen}, {Zambotti}, {Zanoni},
  {Ziegler}, {Zweifel}, {ST7-DRS Operations Team}, {Barela}, {Cutler},
  {Demmons}, {Dunn}, {Girard}, {Hsu}, {Javidnia}, {Li}, {Maghami},
  {Marrese-Reading}, {Mehta}, {O'Donnell}, {Romero-Wolf}, \&
  {Ziemer}}]{thoslubak19}
{Thorpe}, J.~I., {Slutsky}, J., {Baker}, J.~G., {et~al.} 2019, \apj, 883, 53,
  \dodoi{10.3847/1538-4357/ab3649}

\bibitem[{{Watanabe} {et~al.}(2019){Watanabe}, {Hirabayashi}, {Hirata},
  {Hirata}, {Noguchi}, {Shimaki}, {Ikeda}, {Tatsumi}, {Yoshikawa}, {Kikuchi},
  {Yabuta}, {Nakamura}, {Tachibana}, {Ishihara}, {Morota}, {Kitazato},
  {Sakatani}, {Matsumoto}, {Wada}, {Senshu}, {Honda}, {Michikami}, {Takeuchi},
  {Kouyama}, {Honda}, {Kameda}, {Fuse}, {Miyamoto}, {Komatsu}, {Sugita},
  {Okada}, {Namiki}, {Arakawa}, {Ishiguro}, {Abe}, {Gaskell}, {Palmer},
  {Barnouin}, {Michel}, {French}, {McMahon}, {Scheeres}, {Abell}, {Yamamoto},
  {Tanaka}, {Shirai}, {Matsuoka}, {Yamada}, {Yokota}, {Suzuki}, {Yoshioka},
  {Cho}, {Tanaka}, {Nishikawa}, {Sugiyama}, {Kikuchi}, {Hemmi}, {Yamaguchi},
  {Ogawa}, {Ono}, {Mimasu}, {Yoshikawa}, {Takahashi}, {Takei}, {Fujii},
  {Hirose}, {Iwata}, {Hayakawa}, {Hosoda}, {Mori}, {Sawada}, {Shimada},
  {Soldini}, {Yano}, {Tsukizaki}, {Ozaki}, {Iijima}, {Ogawa}, {Fujimoto}, {Ho},
  {Moussi}, {Jaumann}, {Bibring}, {Krause}, {Terui}, {Saiki}, {Nakazawa}, \&
  {Tsuda}}]{wathirhir19}
{Watanabe}, S., {Hirabayashi}, M., {Hirata}, N., {et~al.} 2019, Science, 364,
  268, \dodoi{10.1126/science.aav8032}

\bibitem[{{Weidenschilling} \& {Jackson}(1993)}]{weijac93}
{Weidenschilling}, S.~J., \& {Jackson}, A.~A. 1993, Icarus, 104, 244

\bibitem[{{Weryk} \& {Brown}(2004)}]{werbro04}
{Weryk}, R.~J., \& {Brown}, P. 2004, Earth Moon and Planets, 95, 221,
  \dodoi{10.1007/s11038-005-9034-x}

\bibitem[{{Whipple}(1951)}]{whi51}
{Whipple}, F.~L. 1951, ApJ, 113, 464

\bibitem[{Wisdom \& Holman(1991)}]{wishol91}
Wisdom, J., \& Holman, M. 1991, AJ, 102, 1528

\end{thebibliography}



\end{document}